\documentclass[aps,reprint,superscriptaddress,longbibliography]{revtex4-1}

\pdfoutput=1

\usepackage{graphicx}
\usepackage{amsmath}
\usepackage[dvipsnames]{xcolor}
\usepackage[unicode=true, colorlinks=true, citecolor={blue!80!black}, urlcolor={blue!50!black}, linkcolor = {blue!80!black}]{hyperref}
\usepackage{microtype}
\usepackage{times}

\DeclareMathOperator{\sgn}{sgn}
\DeclareMathOperator{\re}{\mathrm{Re}}
\DeclareMathOperator{\im}{\mathrm{Im}}

\newcommand{\abs}[1]{\left\vert#1\right\vert} 
\newcommand{\pd}{\partial}
\newcommand{\mc}[1]{\mathcal{#1}}
\newcommand{\up}{\uparrow}
\newcommand{\down}{\downarrow}

\begin{document}

\title{Energy spectrum and current-phase relation of a nanowire Josephson junction close to the topological transition}

\author{Chaitanya Murthy}
\affiliation{Department of Physics, University of California, Santa Barbara, CA 93106, USA}

\author{Vladislav D.~Kurilovich}
\affiliation{Department of Physics, Yale University, New Haven, CT 06520, USA}

\author{Pavel D.~Kurilovich}
\affiliation{Department of Physics, Yale University, New Haven, CT 06520, USA}

\author{Bernard van Heck}
\affiliation{Microsoft Quantum, Station Q, University of California, Santa Barbara, CA 93106, USA}
\affiliation{Microsoft Quantum Lab Delft, Delft University of Technology, 2600 GA Delft, The Netherlands}

\author{Leonid I.~Glazman}
\affiliation{Department of Physics, Yale University, New Haven, CT 06520, USA}

\author{Chetan Nayak}
\affiliation{Department of Physics, University of California, Santa Barbara, CA 93106, USA}
\affiliation{Microsoft Quantum, Station Q, University of California, Santa Barbara, CA 93106, USA}

\date{\today}

\begin{abstract}

A semiconducting nanowire proximitized by an $s$-wave superconductor can be tuned into a topological state by an applied magnetic field.
This quantum phase transition is marked by the emergence of Majorana zero modes at the ends of the wire.
The fusion of Majorana modes at a junction between two nanowires results in a $4\pi$-periodic Josephson effect.
We elucidate how the $4\pi$-periodicity arises across the topological phase transition in a highly-transparent short nanowire junction.
Owing to a high transmission coefficient, Majorana zero modes coming from different wires are strongly coupled, with an energy scale set by the proximity-induced, field-independent pairing potential.
At the same time, the topological spectral gap---defined by competition between superconducting correlations and Zeeman splitting---becomes narrow in the vicinity of the transition point.
The resulting hybridization of the fused Majorana states with the spectral continuum strongly affects the electron density of states at the junction and its Josephson energy.
We study the manifestations of this hybridization in the energy spectrum and phase dependence of the Josephson current.
We pinpoint the experimentally observable signatures of the topological phase transition, focusing on junctions with weak backscattering.\\

\end{abstract}

\maketitle

\section{Introduction}

Single-mode semiconductor nanowires proximitized by a conventional $s$-wave superconductor have emerged as a leading candidate for the implementation of topological quantum computing \cite{kitaev2001, lutchyn2010, oreg2010, alicea2012}.
Due to the spin-orbit coupling in the wire, the two Kramers doublets at the Fermi energy, $\pm k_{\rm out}$ and $\pm k_{\rm in}$, respectively, are separated in momentum space.
The superconducting proximity effect acts individually on each of the two doublets, inducing superconducting pairing gaps.
The resulting state is topologically trivial.
In a magnetic field, however, pairing competes with spin polarization induced by the Zeeman effect.
A non-trivial topological state is formed if the Zeeman effect wins for one of the doublets.
For definiteness, we concentrate on the most favorable point for the formation of a topologically non-trivial state, $k_{\rm in}=0$, achievable at a specific value of the Fermi energy.
We also assume that the spin-orbit coupling is strong.
In this case, a homogeneous magnetic field parallel to the wire induces a Zeeman splitting of the $k_{\rm in}=0$ ``inner'' doublet, while having little effect on the $\pm k_{\rm out}$ ``outer'' doublet.
Under these conditions, only the Cooper pairs belonging to the $\pm k_{\rm out}$ helical modes remain intact, giving rise to a topological superconducting state which supports a Majorana zero mode (MZM) at each end of the wire.

Quantum computing operations require controllable fusion of pairs of MZMs belonging to different wires (or to different proximitized portions of the same wire) into a single Dirac
fermion \cite{alicea2011, aasen2016}.
The energy of the resulting fused state is not fixed at zero and depends on the strength and phase $\phi$ of the Josephson coupling between the wires.
If each wire carried only the $\pm k_{\rm out}$ helical mode, this dependence would consist of a single $4\pi$ harmonic with amplitude proportional to the transmission amplitude, $\sqrt{D}$, of the junction \cite{fu2009}.
The fused state would be localized at the junction.

The peculiarity of the topological phase transition in a proximitized nanowire is that the natural energy scale of the coupling between the MZMs remains large (of the order of the proximity-induced pairing potential) even while the spectral gap closes at the transition.
This occurs because the MZMs are formed out of the $\pm k_{\rm out}$ helical states, while the smallest gap in the spectral continuum lies within the states adjacent to the $k_{\rm in}=0$ momentum.
The presence of a continuum with a narrow gap strongly modifies the Dirac fermion formed by fusion at the junction. This modification affects the density of states at the junction and results in an unsual current-phase relation of the Josephson effect.

In this work, we investigate the energy spectrum and zero-temperature thermodynamic properties of the junction as it is tuned through the topological phase transition.
We focus on the experimentally important case of short, almost-transparent junctions ($1-D\ll 1$).
Due to the emphasis on the topological phase transition itself, our work complements previous studies of the $4\pi$-periodic Josephson effect in proximitized nanowires \cite{lutchyn2010,meng2012,pikulin2012,sanjose2013,cayao2015,marra2016,peng2016,nesterov2016,cayao2017}, and provides guidance for experiments aimed at detecting the onset of the topological phase.

\subsection*{Summary of results}

In the absence of backscattering ($D=1$), a junction does not mix the 
states near $k_{\rm in}$ with those near $\pm k_{\rm out}$, which we will refer to as the inner and outer modes respectively, at any $\phi$.
If the induced gap $\Delta$ exceeds the Zeeman energy $B$ for the inner modes, the wires are in a topologically trivial state.
At finite $\phi$, both inner and outer modes carry an Andreev bound state localized at the junction.
These states cross zero energy and simultaneously change their ground state occupation once the phase crosses $\phi=\pm\pi$
(due to periodicity, it is sufficient to consider an interval $\phi\in [-2\pi,2\pi]$).
Such a change of the occupation is allowed, as it does not violate 
fermion parity conservation.
The discontinuous change in the ground state leads to discontinuities of the Josephson current $I(\phi)$ at $\phi=\pm\pi$.
At the critical value of the magnetic field, $B=\Delta$, the gap in the inner modes closes and then reopens at $B>\Delta$ without the bound state.
The remaining bound state (associated with the outer modes) does cross zero energy at $\phi=\pm\pi$, but this time its occupation cannot change, due to fermion parity conservation.
The occupation of the bound state may change only at larger phases, $\phi=\pm(\pi+\phi_0)$, at which placing a quasiparticle in the continuum above the gap in the inner-mode spectrum is energetically favorable.
The peculiarity of the resulting ground state is that it contains a quasiparticle at the edge of the continuum, and therefore is not separated by a gap from the excited states.

The appearance of $\phi_0\neq 0$ at $B>\Delta$ signals that the properties of the junction are $4\pi$-periodic in the topological phase.
While the change of the periodicity in $\phi$ at $B=\Delta$ is associated with a dramatic qualitative change of the eigenstates, the quantitative changes in observables---such as the Josephson current---are more subtle.
Indeed, the inner-mode spectral gap scales linearly with $|B-\Delta|$ and is small near the transition.
Therefore the discontinuities  in $I(\phi)$ shift from the points $\pm\pi$ by small amounts, $\phi_0\propto(B-\Delta)\Theta(B-\Delta)$.

The effect of weak scattering on the energy spectrum and the Josephson current is different---at the qualitative as well as the quantitative level---on the two sides of the topological transition.
In the topologically trivial state, backscattering couples the bound states of the inner and outer modes.
As a result, the bound state energies are repelled from zero in the vicinity of $\phi=\pm\pi$.
This leads to the smearing of the discontinuity in the Josephson current over a phase interval $\delta\phi_{B < \Delta} \propto \sqrt{1-D}$, similar to the standard case of a short SNS junction \cite{beenakker1991}. In the topological state the zero energy crossings at $\phi = \pm \pi$ are protected by fermion parity conservation and are thus unaffected by the scattering.
At the same time, even weak scattering substantially alters the bound state energy once it approaches the edge of the inner-mode continuum.
Hybridization between the bound state and the continuum states smears the discontinuity in the Josephson current over $\delta \phi_{B > \Delta}  \propto (1-D)^2$.
Thus on the topological side of the transition the smearing is weak compared to that on the trivial side. 
Furthermore, we find another spectroscopic feature of the topological transition: in the vicinity of $\phi=0,\pm 2\pi$, weak scattering peels off shallow Andreev bound states from the continuum.
These states appear only in the topologically non-trivial phase, and can thus serve as an additional signature of the transition.

Throughout the paper, we pay particular attention to the contribution of the continuous part of the spectrum to the Josephson properties.
In the short junction limit, the continuum contribution to the ground state energy and to the Josephson current vanishes at zero magnetic field, as is well known, but it becomes nonzero at finite magnetic field.
This contribution is $2\pi$-periodic in both the trivial and topological phases. Yet, it carries an imprint of the transition in the form of a non-analytic dependence on the magnetic field close to the critical point.

The manuscript is structured as follows.
In Section \ref{sec:model}, we introduce and motivate the model of a proximitized wire used in the rest of the work.
In Section \ref{sec:transparent}, we study the properties of a perfectly transparent junction.
We deal with scattering at the junction in Section~\ref{sec:scat}, paying particular attention to the coupling it induces between the bound states and the continuum in the topological phase.
We conclude in Section \ref{sec:conclusions} with a few remarks about experiments and future directions of research.
Technical calculations are left as Appendices. Throughout the work, we use units with $\hbar = 1$.

\begin{figure}[t]
  \begin{center}
    \includegraphics[width=\columnwidth]{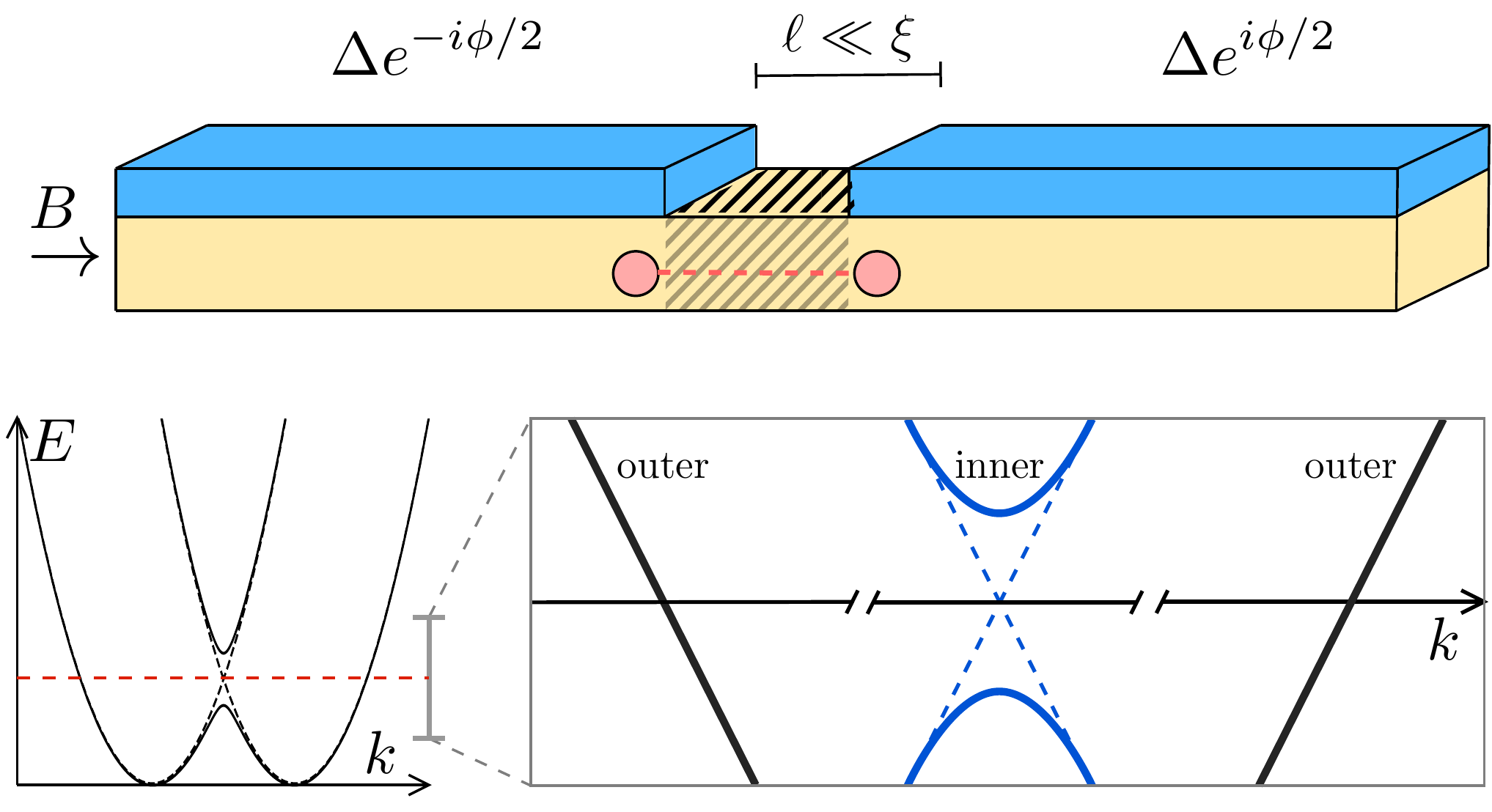}
    \caption{\emph{Top:} Schematic layout of the system under study: a Josephson junction (shaded yellow) formed by a semiconducting wire (yellow) promiximitized by an $s$-wave superconductor (blue). Upon applying a parallel magnetic field, Majorana zero modes may form (red dots) and couple (red dashed line) at the junction. \emph{Bottom:} Bandstructure of a single-subband Rashba wire with an applied magnetic field. If the Fermi level is placed in the middle of the Zeeman gap (dashed red line), the low-energy linearized theory consists of a pair of inner modes and a pair of outer modes. Induced $s$-wave superconducting pairing acts within each pair of modes, while scattering at the junction couples inner and outer modes.}
    \label{fig:wire_layout}
  \end{center}
\end{figure}

\section{The model}
\label{sec:model}

We consider a Josephson junction in a proximitized nanowire with strong Rashba spin-orbit coupling. 
The wire is placed in a magnetic field parallel to its axis.
We assume that the orbital effect of the field can be neglected, and account only for the Zeeman effect of the field. 
We also assume that the junction is short, i.e., its length $\ell$ is much smaller than the superconducting coherence length $\xi$ (see Fig.~\ref{fig:wire_layout} for an illustration of the setup).
To model this system, we start with the mean-field many-body Hamiltonian, which takes the form \cite{oreg2010,lutchyn2010}
\begin{equation}
\label{eq:HBCS}
    \hat{H}=\frac{1}{2}\int dx\, \hat{\Psi}^\dagger(x)\,\mathcal{H}\,\hat{\Psi}(x),
\end{equation}
where $\hat{\Psi} = (\hat{\psi}_\up, \hat{\psi}_\down, \hat{\psi}^\dagger_\down, -\hat{\psi}^\dagger_\up)^T$, $\hat{\psi}_\sigma$ is the annihilation operator of an electron with spin $\sigma=\:\uparrow{\rm or} \downarrow$, and the Bogoliubov-de Gennes (BdG) Hamiltonian is given by
\begin{equation}
\label{eq:HBdG}
    \mathcal{H} = \left[-\frac{\pd_x^2}{2m}-i\alpha\pd_x\,\sigma_z - \mu + V(x)\right]\tau_z - B\,\sigma_x + \Delta(x)\,\tau_x .
\end{equation}
Here $\sigma_i$ and $\tau_i$ are Pauli matrices in spin and Nambu space, respectively; $m$ is the effective mass, $\alpha$ is the spin-orbit coupling constant, $\mu$ is the chemical potential, $B$ is the Zeeman energy (without loss of generality we take $B\geq 0$), and $\Delta(x)$ is the proximity-induced superconducting order parameter:
\begin{equation}
\label{eq:Delta_x}
    \Delta(x) = \Delta \, e^{i(\phi/2)\sgn(x)\,\tau_z} ,
\end{equation}
where $\phi$ is the phase difference across the junction.
The potential $V(x)$ accounts for a barrier that scatters electrons at the junction.
Instead of specifying a particular functional form for $V(x)$, we will, in the following, account for its effect by imposing suitable boundary conditions at $x=0$; these boundary conditions will be formulated directly in terms of the mesoscopic scattering parameters (transmission probability and scattering phases) of the junction in the normal state.

We study the evolution of the junction properties as the Hamiltonian is tuned across the topological phase transition by changing the magnetic field.
In the model defined by Eq.~\eqref{eq:HBdG}, the transition from the topologically trivial ($B<B_\mathrm{c}$) to non-trivial ($B>B_\mathrm{c}$) phase happens at $B_\mathrm{c}=(\Delta^2+\mu^2)^{1/2}$ \cite{oreg2010}.
Throughout this work, we assume, for simplicity, that the chemical potential is set at the optimal point $\mu=0$ where the critical field is minimal, $B_\mathrm{c} = \Delta$.

Following Ref.~\cite{vanheck2017}, we further assume that the spin-orbit coupling is strong, $m\alpha^2\gg \Delta, B$.
The latter condition implies that the low-energy spectrum in the bulk of the nanowire consists of well-separated modes in the vicinity of the Fermi points $k = k_\mathrm{in} = 0$ (the inner modes) and $k = \pm k_\mathrm{out} = \pm 2m\alpha$ (the outer modes).
On the technical level, the condition $m\alpha^2\gg \Delta, B$ allows us to invoke the Andreev approximation and to expand the field operator $\hat{\Psi}$ into helical components involving modes close to the momenta $k_\mathrm{in}$ and $\pm k_\mathrm{out}$ \cite{klinovaja2012,vanheck2017}:
\begin{equation}
\label{eq:decomp}
    \hat{\Psi}(x) = e^{-im\alpha x (1+\sigma_z)} \hat{\Psi}_L(x) 
    + e^{im\alpha x(1-\sigma_z)} \hat{\Psi}_R(x) .
\end{equation}
Here $\hat{\Psi}_L$ and $\hat{\Psi}_R$ denote left- and right-moving components of the field, which both have Fermi velocity given by $\alpha$. 
Then, by inserting the decomposition \eqref{eq:decomp} into Eq.~\eqref{eq:HBCS} and averaging out rapidly oscillating terms, we obtain a set of BdG equations for the inner ($\Phi_\textrm{i}$) and outer ($\Phi_\textrm{o}$) mode wave functions  at $x\neq 0$,
\begin{subequations}
\label{eq:BdG_equations}
\begin{align}
    [-i\alpha \pd_x \tau_z \sigma_z - B \sigma_x + \Delta(x)\tau_x] \Phi_\textrm{i}(x) &= E \Phi_\textrm{i}(x) , 
    \label{eq:BdG_equations_a} \\*
    [+i\alpha \pd_x \tau_z \sigma_z + \Delta(x)\tau_x] \Phi_\textrm{o}(x) &= E \Phi_\textrm{o}(x) . 
    \label{eq:BdG_equations_b}
\end{align}
\end{subequations}
Notice that the Zeeman energy drops out from the BdG equations for the outer modes.
This is a consequence of a large energy separation $\sim m\alpha^2 \gg B$ between spin subbands in the vicinity of $k=\pm k_\mathrm{out} = \pm 2m\alpha$.

As discussed above, Eqs.~\eqref{eq:BdG_equations} should be supplemented by a suitable boundary condition at $x = 0$.
Within the Andreev approximation the boundary condition is completely determined by the scattering matrix of the junction in the normal state ($\Delta = B = 0$).
Under the assumption that the scattering is spin-independent at $\Delta, B = 0$, the most general boundary condition is:
\begin{subequations}
\label{eq:BdG_bc}
\begin{align}
    \Phi_\textrm{i}(0^+) &= \frac{e^{i\gamma \sigma_z}}{\sqrt{D}}\Bigl[ \Phi_\textrm{i}(0^-) 
    + e^{-i\delta \sigma_z}\sqrt{1-D}\, \Phi_\textrm{o}(0^-)\Bigr], 
    \label{eq:BdG_bc_a} \\*[0.2em]
    \Phi_\textrm{o}(0^+) &= \frac{e^{-i\gamma \sigma_z}}{\sqrt{D}} \Bigl[\Phi_\textrm{o}(0^-) 
    + e^{i\delta  \sigma_z} \sqrt{1-D} \,\Phi_\textrm{i}(0^-)\Bigr].
    \label{eq:BdG_bc_b}
\end{align}
\end{subequations}
Here, $D$ is the normal-state transmission probability, $\gamma \in [-\pi / 2, \pi / 2]$ is the forward scattering phase in the normal state, and $\delta$ is the reflection phase in the normal state for a particle incoming from $x = - \infty$.
Notice that the reflection phase $\delta$ can be eliminated from Eqs.~\eqref{eq:BdG_equations}, \eqref{eq:BdG_bc} by a unitary transformation $\Phi_\mathrm{o} \rightarrow e^{i\delta\sigma_z}\Phi_\mathrm{o}$. Therefore, we suppress it in the following discussions. The independence of the properties of the junction on $\delta$ is a consequence of the $m\alpha^2 \gg B,\Delta$ approximation, in which the states in the outer modes are insensitive to the magnetic field. Throughout this work, we assume that $D$ and $\gamma$ are independent of energy up to the relevant scales $|E|\sim B,\Delta$.

For future reference, we mention that in the particular case of a (repulsive) delta function barrier, $V(x) = \kappa \, \delta(x)$, the parameters $D$ and $\gamma$ are given by
\begin{equation}
\label{eq:deltabarrier}
    D = \frac{1}{1+(\kappa/\alpha)^2}, \qquad 
    \gamma = -\arctan\sqrt{\frac{1-D}{D}}.
\end{equation}
However, in general, there is no rigid connection between $D$ and $\gamma$ like the one provided by Eq.~\eqref{eq:deltabarrier}. We treat them as independent parameters in what follows.

In the next section (Sec.~\ref{sec:transparent}) we study the spectrum of the junction and its thermodynamic properties (at $T = 0$) in the limit of perfect transmission, $D=1$ and $\gamma = 0$.
Then in Sec.~\ref{sec:scat} we discuss the effects induced by scattering at the junction, $D < 1$ and $\gamma \neq 0$.

\section{Spectrum and thermodynamic properties of a transparent junction}
\label{sec:transparent}

If $D=1$, the inner and outer modes decouple from each other, as follows from Eq.~\eqref{eq:BdG_bc}. This simplification allows us to find a compact analytical solution of the BdG equations \eqref{eq:BdG_equations} at $\gamma = 0$ via a standard wave function matching procedure. 
The energy spectrum consists of a discrete part formed by Andreev bound states localized at the junction and a continuous part formed by extended states at energies above the gaps in their respective modes.
In the outer modes the gap equals $\Delta$, independent of $B$, while in the inner modes the gap is $\abs{\Delta-B}$; it closes at the topological transition, $B=\Delta$.

\begin{figure*}[t]
  \begin{center}
    \includegraphics[width=0.9\textwidth]{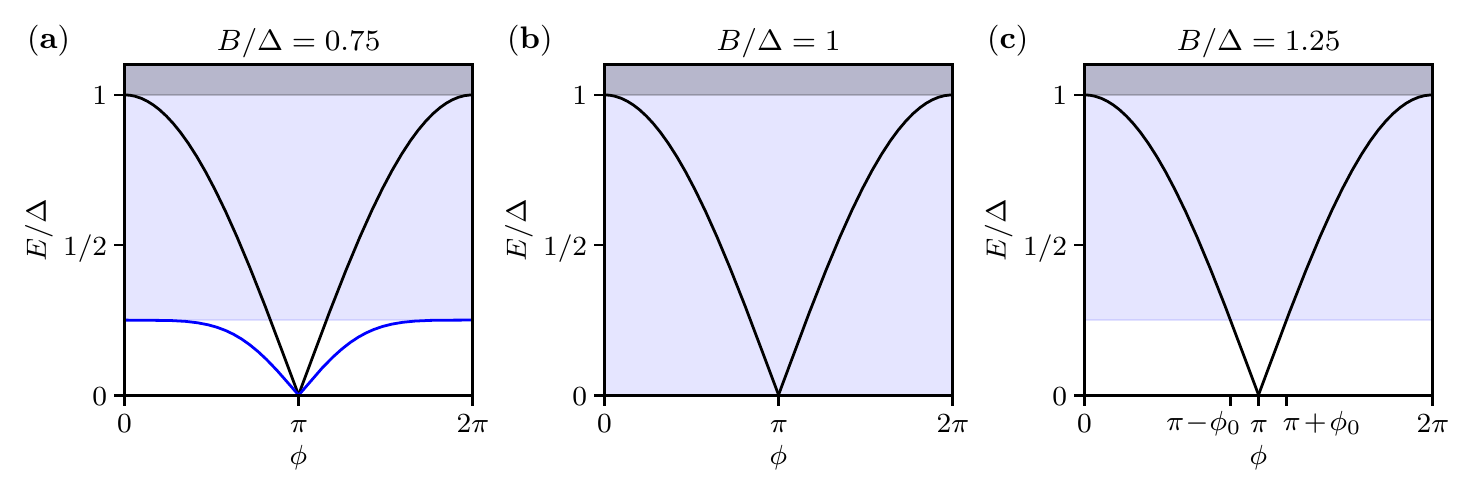}
    \caption{Evolution of the single-particle excitation spectrum of the junction with increasing Zeeman energy, in the limit of perfect transmission, $D = 1$, $\gamma = 0$. Solid curves represent the discrete energy levels. Shaded regions mark the continuous spectrum. The latter consists only of inner modes for $E<\Delta$, and of both inner and outer modes for $E\geq \Delta$. In the trivial phase, $B < \Delta$, there are bound states in the inner and outer modes (black and blue curves in panel (a), respectively). At $B=\Delta$ the gap in the inner modes closes and the corresponding bound state vanishes, see panel (b). In the topological phase, $B > \Delta$, only the outer-mode bound state remains [panel (c)]. As long as $B < 2\Delta$ the energy of the bound state in the outer modes overlaps with the inner-mode continuum in some domain of phase $\phi$.}
    \label{fig:andreev_levels}
  \end{center}
\end{figure*}

\subsection{Bound states at perfect transmission}
\label{sec:bs}

We start by considering the bound state in the outer modes.
In this case, the BdG equations are similar to those of a proximitized helical edge in a quantum spin Hall state \cite{fu2009}. 
Eq.~\eqref{eq:BdG_equations_b} decouples for left-movers ($\sigma_z = +1$) and right-movers ($\sigma_z = -1$). 
By matching the wavefunctions continuously across the junction (as required by Eq.~\eqref{eq:BdG_bc} with $D=1$, $\gamma = 0$), we obtain the equations for the bound state energies in the form $\Lambda_{\mathrm{o}, \sigma}(E,\phi) = 0$, where
\begin{equation}
\label{eq:Lambda_o}
    \Lambda_{\mathrm{o}, \sigma}(E,\phi) = \sin[\sigma \eta(E) - \phi/2],
\end{equation}
with $e^{i \eta(E)} = E/\Delta + i \sqrt{1 - (E/\Delta)^2}$ and $\sigma = \pm 1$ defining the eigenvalue of $\sigma_z$. Eq.~\eqref{eq:Lambda_o} results in the \emph{non-degenerate} bound state solution $E=\pm E_0(\phi)$ with
\begin{equation}
\label{eq:E0}
    E_0(\phi) = \Delta \cos(\phi/2),
\end{equation}
independent of $B$.

For the inner modes, solving the BdG equations is more cumbersome due to the presence of the Zeeman term in Eq.~\eqref{eq:BdG_equations_a}. Bound states may appear at $\abs{E}<\abs{\Delta-B}$.
The bound state energy is a solution of the equation $\Lambda_{\mathrm{i}}(E,\phi) = 0$, with
(see Appendix~\ref{app:imode_bound_state})
\begin{align}
\label{eq:Lambda_i}
    \Lambda_\mathrm{i} (E,\phi) = \, &[(\Delta-B)^2-E^2]^{1/2}[(\Delta+B)^2-E^2]^{1/2} \nonumber \\* 
    &- (E^2+\Delta^2-B^2) F(\phi) .
\end{align}
The phase dependence of this expression is encoded in the function
\begin{equation}
\label{eq:F_def}
    F(\phi) \equiv \frac{\sin^2(\phi/2)}{\cos^2(\phi/2) + 1} .
\end{equation}

The equation $\Lambda_\mathrm{i}(E,\phi) = 0$ admits solutions only in the topologically trivial phase, $B<\Delta$.
Indeed, at $B > \Delta$  and $|E|<|\Delta - B|$, the quantity $(E^2 + \Delta^2 - B^2)$ in the second line of Eq.~(\ref{eq:Lambda_i}) is negative, and hence $\Lambda_{\mathrm{i}}(E,\phi) > 0$.
The bound state solution at $B<\Delta$ has energy $E=\pm E_1(\phi)$ with
\begin{equation}
\label{eq:bs_energy}
    E_1(\phi) = \frac{\Delta-\sqrt{\Delta^2-(\Delta^2-B^2)[1-F^2(\phi)]}}{\sgn[\cos(\phi /2)] \sqrt{1-F^2(\phi)}}\,.
\end{equation}
This expression has several notable features.
First, $E_1(\phi)=E_0(\phi)$ at $B=0$, i.e., the Andreev spectrum is two-fold degenerate in the absence of the magnetic field.
Second, $E_1(\phi)$ crosses zero at $\phi = \pi$, simultaneously with the energy of the outer-mode bound state $E_0(\phi)$.
The presence of such a crossing is a peculiarity of the perfect transmission limit, which is not robust to the presence of backscattering.
Third, the energy $\abs{E_1(\phi)}$ is bounded by $\Delta-B$, so its phase dispersion is suppressed by the magnetic field, until the bound state merges with the continuum at $B=\Delta$, in concurrence with the topological phase transition.

In the topological phase, $B > \Delta$, the bound state is no longer present in the inner modes; only the outer-mode bound state, which can be considered as two fused Majorana modes, remains at the junction.
For $B<2\Delta$ the energy of the latter, $E_0$, overlaps with the inner-mode continuum, intersecting its edge at $\phi=\pi \pm \phi_0$, where
\begin{equation}
\label{eq:phi0}
    \phi_0 = 2 \arcsin\!\left(\frac{B-\Delta}{\Delta}\right).
\end{equation}
This coexistence of a bound state with the continuous part of the spectrum is another peculiarity of the perfect transmission limit. 
The evolution of the Andreev spectrum upon increasing Zeeman energy is shown in Fig.~\ref{fig:andreev_levels}.

To conclude this section, we note that for $B>2\Delta$ the gap in the inner modes, $B - \Delta$, exceeds the gap in the outer modes, $\Delta$.
Consequently, the bound state energy $E_0$ is separated from the continuum at all phases (except at the points $\phi=0, 2\pi$).
In this high-magnetic field topological regime, the low-energy spectrum of the model becomes identical to that of the Fu-Kane model of a proximitized quantum spin Hall edge.
Then, on a qualitative level the properties of the junction at $B>2\Delta$ are similar to those described in Ref.~\cite{fu2009}.
In what follows we concentrate instead on the vicinity of the topological transition, $B < 2\Delta$.

\subsection{Continuum states at perfect transmission}
\label{sec:cont}

The continuous part of the Andreev spectrum consists of scattering states that appear at energies above the threshold values $\abs{\Delta - B}$, $\Delta$, $\Delta + B$.
The total density of continuum states at energy $E$, $\rho(E,\phi)$, can be represented in the form
\begin{equation} \label{eq:dos}
    \rho(E,\phi) = L\,g(E) + \delta \rho (E,\phi).
\end{equation}
Here $L$ is the system size, $g(E)$ is the bulk density of states per unit length, and $\delta \rho(E,\phi)$ is the correction to the density of states due to the presence of the junction.
$g(E)$ is given by a sum of three BCS-like terms with gap parameters determined by the threshold energies:
\begin{equation}
    g(E) = g_{\abs{\Delta - B}}(E) + 2 g_{\Delta}(E) + g_{\Delta +B}(E) ,
\end{equation}
where
\begin{equation}
    g_{\varepsilon}(E) \equiv \Theta(E -\varepsilon) \, \frac{E}{\pi \sqrt{E^2 - \varepsilon^2}}
\end{equation}
and $\Theta(x)$ is the Heaviside step function. The phase-dependent contribution $\delta \rho(E,\phi)$ can be recovered from the quasiparticle scattering matrix of the junction, $S(E,\phi)$, via the relation \cite{akkermans1991, souma2002}
\begin{equation}
\label{eq:rho_S}
    \delta\rho(E,\phi) = \frac{1}{2\pi i} \frac{\partial}{\partial E} \ln \det S(E,\phi).
\end{equation}
We note that at finite $\phi$ the scattering matrix is non-trivial even in the perfect transmission limit $D=1,\,\gamma = 0$.
For $|\Delta - B| < E < \Delta$ the scattering states reside in the inner modes only.
In this case, the determinant of the scattering matrix $S \equiv S_\mathrm{i}$ is given by (see Appendix~\ref{app:imode_s_matrix} for details)
\begin{equation}
\label{eq:det_Si}
    \det S_\mathrm{i}(E,\phi) = \frac{[\Lambda_\mathrm{i}(E,\phi)]^\star}{\Lambda_\mathrm{i}(E,\phi)},
\end{equation}
where ${}^\star$ denotes complex conjugation, and where the function $\Lambda_{\mathrm{i}}$, defined in Eq.~\eqref{eq:Lambda_i}, should be analytically continued from the range $|E| < |\Delta - B|$ into the interval $E \in (|\Delta - B|,\Delta)$ by taking $[(\Delta-B)^2 - E^2]^{1/2} \to -i [E^2 - (\Delta-B)^2]^{1/2}$. 

For larger energies, $E > \Delta$, there are scattering states both in the inner and in the outer modes.
Owing to perfect transmission, the scattering matrix is block-diagonal in these subspaces.
Therefore, its determinant is multiplicative, $\det S = (\det S_\mathrm{i}) \cdot (\det S_{\mathrm{o},+}) \cdot (\det S_{\mathrm{o},-})$, where $\det S_\mathrm{i}$ is given by Eq.~\eqref{eq:det_Si} and $\det S_{\mathrm{o},\sigma}$ by
\begin{equation}
\label{eq:det_So}
    \det S_{\mathrm{o},\sigma}(E,\phi) = \frac{[\Lambda_{\mathrm{o},\sigma}(E,\phi)]^\star}{\Lambda_{\mathrm{o},\sigma}(E,\phi)}.
\end{equation}
From Eq.~\eqref{eq:Lambda_o} it follows that $\Lambda_{\mathrm{o},+} = (\Lambda_{\mathrm{o},-})^\star$ at $E > \Delta$.
Therefore, $(\det S_{\mathrm{o},+}) \cdot (\det S_{\mathrm{o},-}) = 1$ and the outer-mode continuum states do not contribute to $\delta\rho(E,\phi)$.
As a result, at all energies $E > |\Delta - B|$ the density of states correction is given by $\delta \rho (E,\phi) = (2\pi i)^{-1} \partial_E \ln \det S_\mathrm{i}$.
Explicit calculation for $|\Delta - B| < E < \Delta + B$ yields
\begin{equation}
\label{eq:rho_i}
    \delta \rho = \frac{\partial}{\partial E} \frac{1}{\pi} 
    \arctan \frac{(E^2 + \Delta^2 - B^2) F(\phi)}{\sqrt{[E^2-(\Delta-B)^2][(\Delta+B)^2-E^2]}} ,
\end{equation}
where $F(\phi)$ is defined in Eq.~(\ref{eq:F_def}).
The function $\Lambda_\mathrm{i}(E,\phi)$ is real at $E > \Delta + B$  [see Eq.~\eqref{eq:Lambda_i}], thus $\det S_{\mathrm{i}}(E,\phi) = 1$ and $\delta \rho (E,\phi) = 0$ in that energy domain.

At zero magnetic field, the phase-dependent correction $\delta \rho(E,\phi)$ vanishes at all energies above the continuum edge, a well-known result for a short SNS junction with time-reversal symmetry \cite{beenakker1991}. At nonzero magnetic field, on the other hand, the correction is finite and it carries a signature of the topological phase transition, as will be discussed in more detail below.

\begin{figure*}[t]
  \begin{center}
    \includegraphics[width=0.9\textwidth]{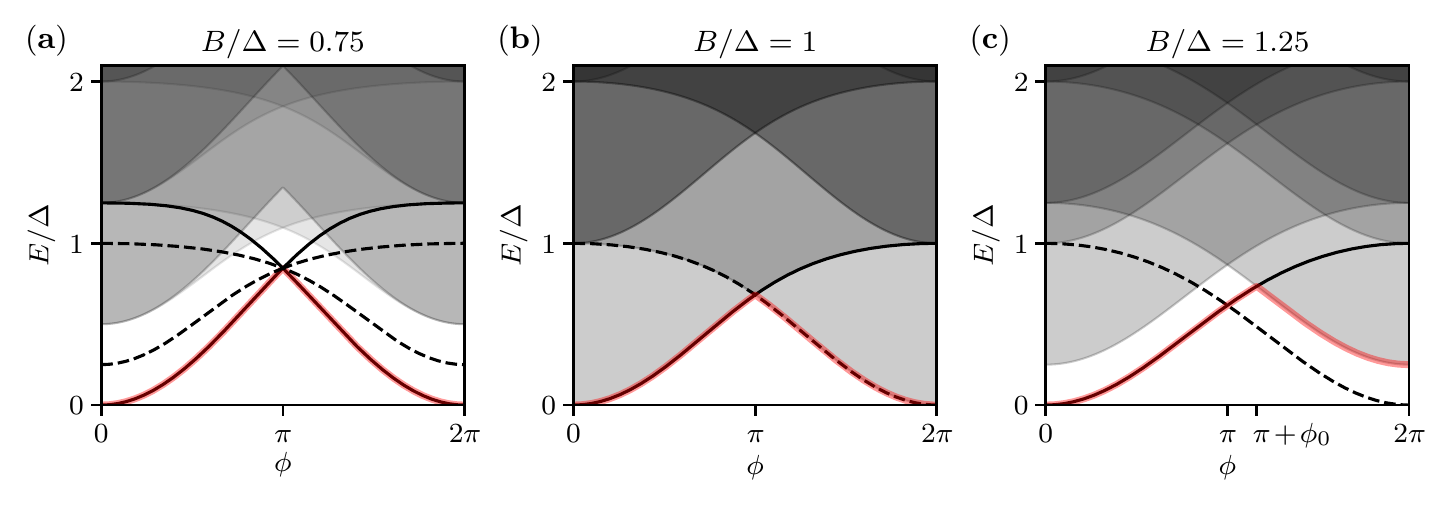}
    \caption{Evolution of the many-body spectrum with increasing Zeeman energy, in the limit of perfect transmission, $D = 1$, $\gamma = 0$. The panels are in one-to-one correspondence with the panels of Fig.~\ref{fig:andreev_levels}. Solid (dashed) curves represent many-body states with even (odd) fermion parity belonging to the discrete spectrum. Shaded grey regions represent the continuous part of the spectrum, of either fermion parity. The solid red curves denote the ground state energy in the even parity sector. Note that the phase dispersion of the ground state energy includes the contribution coming from the continuous part of the BdG spectrum.}
    \label{fig:manybodyspectrum}
  \end{center}
\end{figure*}

\subsection{Thermodynamic properties of the transparent junction}

The detailed description of the Andreev bound states and continuum states in the nanowire junction setup, provided in Secs.~\ref{sec:bs} and \ref{sec:cont}, sets the stage for the discussion of thermodynamic properties of the system.
In this section we calculate the ground state energy of the junction, $E_\mathrm{gs}(\phi)$ (Sec.~\ref{sec:GS_transparent}), and with its help establish the current-phase relation $I(\phi)$ of the Josephson effect at $T = 0$ (Sec.~\ref{sec:Iphi_transparent}).
In view of the periodicity of these properties, hereafter we constrain the phase difference across the junction to the interval $\phi \in [-2\pi, 2\pi]$.

\subsubsection{Ground state structure and many-body spectrum}
\label{sec:GS_transparent}

We start by investigating the structure of the many-body ground state of the system.
To this end we express the many-body Hamiltonian \eqref{eq:HBCS} in its eigenbasis:
\begin{equation}
\label{eq:mbH-2}
    \hat{H} = \sum_\mathrm{b} E_\mathrm{b}(\phi) \bigl(\hat{n}_\mathrm{b} - \tfrac{1}{2} \bigr) 
    + \sum_\mathrm{c} E_\mathrm{c}(\phi) \bigl(\hat{n}_\mathrm{c} - \tfrac{1}{2} \bigr) + \cdots .
\end{equation}
Here the sum in the first term runs over the bound states. In the trivial phase, $B < \Delta$, the index $\mathrm{b} = 0, 1$, with $E_0(\phi)$ and $E_1(\phi)$ given by Eqs.~\eqref{eq:E0} and \eqref{eq:bs_energy}, respectively. In the topological phase, $B > \Delta$, a bound state is present in the outer modes only and $\mathrm{b} = 0$. $\hat{n}_{\mathrm{b}}$ are the occupation number operators of the corresponding Dirac fermions.
The sum in the second term covers all continuum states with energies above the gap, $E_\mathrm{c} > |\Delta - B|$; $\hat{n}_\mathrm{c}$ are the number operators for the fermions in these states.
Finally, the dots stand for a phase-independent additive term in the energy.

To find the ground state energy $E_\mathrm{gs}(\phi)$ we minimize the  Hamiltonian \eqref{eq:mbH-2} 
under the constraint of a fixed fermion parity.
Up to a phase-independent constant, $E_\mathrm{gs}(\phi)$ can be conveniently divided into two parts:
\begin{equation}
\label{eq:EgsEgs1Egs2}
    E_\mathrm{gs}(\phi) = E^{(1)}_\mathrm{gs}(\phi) + E^{(2)}_\mathrm{gs}(\phi).
\end{equation}
The first part, $E^{(1)}_\mathrm{gs}(\phi)$, incorporates the contributions to the ground state energy from the bound states and from the term 
$\sum_\mathrm{c} E_\mathrm{c} \hat{n}_c$ in the Hamiltonian \eqref{eq:mbH-2}. 
The second part, $E^{(2)}_\mathrm{gs}(\phi)$, is a residual contribution due to the quasiparticle continuum arising from the $c$-number term $-\frac{1}{2}\sum_\mathrm{c} E_\mathrm{c}$.
By introducing the correction to the continuum density of states $\delta\rho(E,\phi)$ (see Sec.~\ref{sec:cont}), we represent $E^{(2)}_\mathrm{gs}(\phi)$ in the form
\begin{equation}
\label{eq:cont}
    E^{(2)}_\mathrm{gs}(\phi) = -\frac{1}{2} \int_{\abs{\Delta - B}}^{+\infty} dE E \left[\delta\rho(E,\phi) - \delta\rho(E,0^+)\right] ,
\end{equation}
where we added a constant shift to ensure that $E^{(2)}_\mathrm{gs}\to0$ for $\phi\to 0$.

We begin the investigation of $E_\mathrm{gs}(\phi)$ by considering a junction in a topologically trivial state, $B < \Delta$. For concreteness, we focus on the even fermion parity sector.
As a first step, we establish the structure of the ground state in terms of occupation numbers at different phases $\phi$. 
In the interval $\phi \in [-\pi, \pi]$ neither of the bound states is occupied in the ground state.
At $\phi = \pm \pi$ the energies $E_0$ and $E_1$ of Eq.~\eqref{eq:mbH-2} simultaneously cross zero.
Then, for larger phases $\phi \in (\pi,2\pi]$ and $\phi \in [-2\pi,-\pi)$ it is energetically favorable for two electrons of a Cooper pair to occupy the two bound states.
Such a redistribution is allowed by fermion parity conservation and results in $2\pi$-periodicity of the thermodynamic properties of the system.

An explicit expression for $E_\mathrm{gs}^{(1)}$ at $B < \Delta$ reads
\begin{gather}
\label{eq:Egs_triv_1}
    E_\mathrm{gs}^{(1)}(\phi) = \tfrac{1}{2} \left(\Delta-\left|E_0(\phi)\right|\right) + \tfrac{1}{2}\left(\Delta-B-\left|E_{1}(\phi)\right|\right),
\end{gather}
where the bound state energies $E_{0,1}$ are given by Eqs.~\eqref{eq:E0} and \eqref{eq:bs_energy}.
The absolute values in Eq.~\eqref{eq:Egs_triv_1} are associated with the changes in the occupation numbers of the bound states at $\phi = \pm \pi$.
For convenience, we added a constant offset in Eq.~\eqref{eq:Egs_triv_1} such that $E_\mathrm{gs}^{(1)}(0) = 0$. 

Next, we discuss the continuum contribution to the ground state energy, $E_\mathrm{gs}^{(2)}$.
It may be found by performing the integration in Eq.~\eqref{eq:cont} with $\delta\rho(E,\phi)$ given by Eq.~\eqref{eq:rho_i}.
When the magnetic field is tuned to the vicinity of the topological phase transition, the integral can be evaluated analytically.
For $\Delta - B \ll \Delta$, we find:
\begin{align}
\label{eq:cont_vic}
    E_\mathrm{gs}^{(2)}(\phi) \approx \ &\frac{\Delta}{2}\left[1-\frac{2}{\pi} \frac{\arccos F(\phi)}{\sqrt{1-F^{2}(\phi)}}\right] \nonumber \\*
    &+ \frac{F(\phi)}{2\pi}\left(\Delta - B\right)\ln\frac{\Delta}{|\Delta-B|}, 
\end{align}
with $F(\phi)$ defined in Eq.~\eqref{eq:F_def}. 
Thus, there is critical behavior in $E_\mathrm{gs}$ close to the topological transition: when $B$ approaches $\Delta$, the ground state energy behaves non-analytically as a function of the difference $(\Delta-B)$; see the last term in Eq.~\eqref{eq:cont_vic}. 
In principle, such non-analytic behavior may be probed experimentally, for instance in the dependence of the Josephson plasma frequency on the magnetic field, and may serve as an additional signature of the topological phase transition.

In addition to the ground state energy, Eq.~\eqref{eq:mbH-2} contains information about the spectrum of excited states.
The many-body spectrum of the transparent junction in the topologically trivial phase is shown in Fig.~\ref{fig:manybodyspectrum}(a).
Notice that for $B < \Delta$ there is always an energy gap between the ground state and the quasiparticle continuum.

\begin{figure*}[t]
  \begin{center}
    \includegraphics[width=0.85\textwidth]{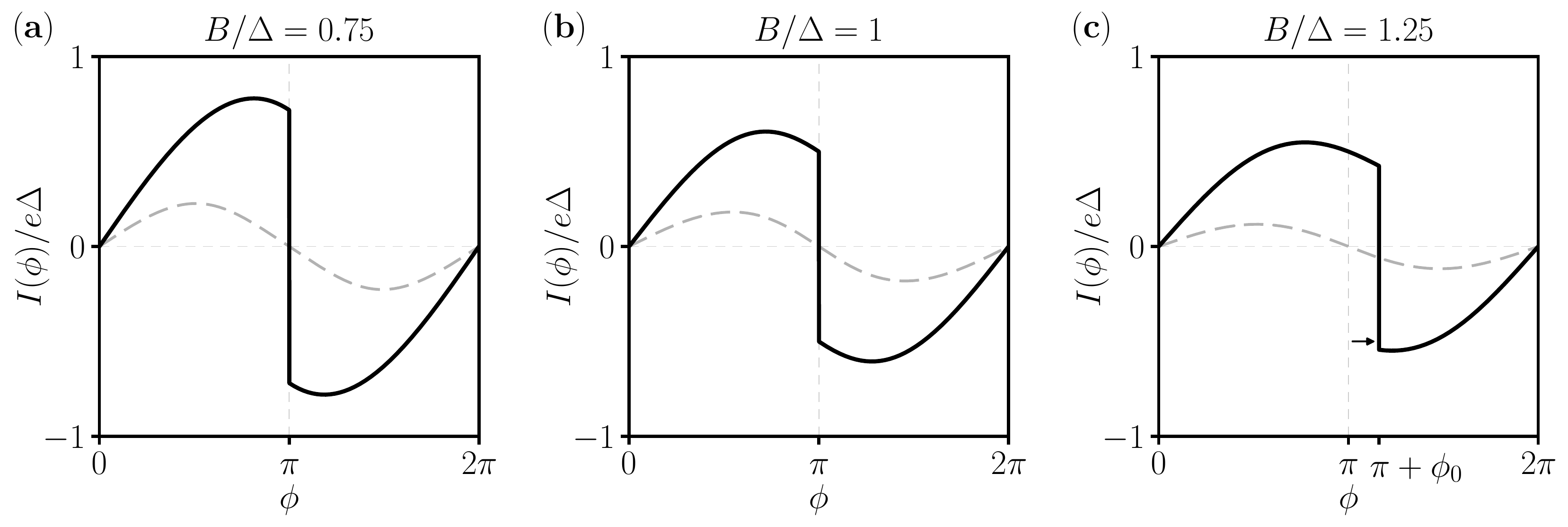}
    \caption{Evolution of the Josephson current $I(\phi)$ with increasing Zeeman energy in the limit of perfect transmission, $D = 1$, $\gamma = 0$ (solid black curves). 
    The panels are in one-to-one correspondence with the panels of Fig.~\ref{fig:manybodyspectrum}.
    In the trivial regime [$B < \Delta$, panel (a)] there is a discontinuity in the current at $\phi = \pi$. Its magnitude is determined by Eq.~\eqref{eq:disc_triv}.  Precisely at the topological transition [$B = \Delta$, panel (b)], the jump in the current decreases to $\delta I_{B = \Delta} = e\Delta$. 
    In the topological state [$B > \Delta$, panel (c)], the discontinuity shifts from $\phi = \pi$ to $\phi = \pi + \phi_0$ [$\phi_0$ is given by Eq.~\eqref{eq:phi0}], signifying the onset of the $4\pi$-periodic Josephson effect. The magnitude of the discontinuity further decreases in accordance with Eq.~\eqref{eq:disc_top}. In each panel, a dashed gray curve depicts separately the continuum contribution, $I^{(2)}(\phi)$, to the Josephson current $I(\phi)$. This contribution is always $2\pi$-periodic and smooth.}
    \label{fig:josephson}
  \end{center}
\end{figure*}

The structure of the ground state in the topological regime, $B>\Delta$, is radically different from that in the trivial state.
At $B>\Delta$ there is only one Andreev bound state, which comes from the outer modes.
At small phase differences $\phi$, this state is not filled.
When the phase reaches $\phi = \pm \pi$ the bound state energy $E_0$ crosses zero, but the level occupation cannot change as that would violate fermion parity conservation.
At $\phi = \pm(\pi + \phi_0)$, where $\phi_0 = 2\arcsin((B-\Delta)/\Delta)$ [see Eq.~\eqref{eq:phi0}], the outer-mode bound state crosses the edge of the inner-mode continuum and, in terms of energy minimization, it becomes profitable to simultaneously occupy the Andreev state and a single quasiparticle state in the continuum.
Therefore, for $\phi\in (\pi + \phi_0, 2\pi]$ and $\phi \in [-2\pi, -(\pi+\phi_0))$, there is no gap in the many-body spectrum between the ground state and the quasiparticle continuum, as depicted in Fig.~\ref{fig:manybodyspectrum}(c).
This feature is in a sharp contrast to the case of the topologically trivial junction, where a gap is present at all values of the phase difference.

The structure of the ground state implies that in the topological state $E_\mathrm{gs}^{(1)}$ is given by
\begin{multline}
\label{eq:Egs_top_1}
    E^\mathrm{(1)}_\mathrm{gs}(\phi) = 
    \frac{\Delta}{2} \Bigl(1-\cos\frac{\phi}{2}\Bigr) \\
    + \Bigl[ \Delta\cos\frac{\phi}{2} + (B-\Delta) \Bigr] \Theta\Bigl(\cos\frac{\pi + \phi_0}{2} - \cos\frac{\phi}{2}\Bigr). \!\!
\end{multline}
Here the second line describes the simultaneous occupation of the Andreev bound state and a state at the edge of the continuum, $E_\mathrm{c}=B-\Delta$, occurring at $\phi = \pm (\pi + \phi_0)$. Expression \eqref{eq:Egs_top_1} is manifestly $4\pi$-periodic, which indicates the onset of the fractional Josephson effect on the topological side of the transition.

In contrast to $E^\mathrm{(1)}_\mathrm{gs}(\phi)$, the continuum contribution to the ground state energy, $E_\mathrm{gs}^{(2)}$, is $2\pi$-periodic even in the topological phase [as follows from Eq.~\eqref{eq:rho_i}].
Close to the transition threshold, $B - \Delta \ll \Delta$, $E_\mathrm{gs}^{(2)}$ is described by Eq.~\eqref{eq:cont_vic}, similarly to the case $B<\Delta$.
Therefore, the logarithmic behavior in Eq.~\eqref{eq:cont_vic} is characteristic to both sides of the topological transition.

\subsubsection{Josephson current}
\label{sec:Iphi_transparent}

Having discussed the ground state structure and $E_\mathrm{gs}(\phi)$ on both sides of the topological transition, we proceed to the evaluation of the Josephson current, $I(\phi)$. At zero temperature, $I(\phi)$ is related to the ground state energy via
\begin{equation}
\label{eq:JC}
    I(\phi) = 2e \, \frac{dE_\mathrm{gs}(\phi)}{d\phi} ,
\end{equation}
where $e$ is the electron charge, and thus can be calculated at arbitrary ratio $B/\Delta$  by employing the results of Sec.~\ref{sec:GS_transparent}.

First, we compute $I(\phi)$ in the topologically trivial state, $B < \Delta$. 
An example of the current-phase relation obtained from Eqs.~(\ref{eq:EgsEgs1Egs2}--\ref{eq:Egs_triv_1}) and \eqref{eq:JC} is presented in Fig.~\ref{fig:josephson}(a). A notable feature of the resulting $I(\phi)$ is the presence of a discontinuity. It occurs at $\phi=\pi$ (and, similarly, at $\phi = -\pi$) and originates from switching in the occupation numbers of the Andreev bound states. In general, such stepwise behavior of the Josephson current is common for short transparent junctions. A peculiarity of the nanowire setup is the dependence of the magnitude $\delta I$ of the discontinuities on the magnetic field. Using Eq.~\eqref{eq:Egs_triv_1} we find that it is given by
\begin{equation}
\label{eq:disc_triv}
    \delta I_{B<\Delta} = e\Delta \left(1 + \frac{\Delta^2 - B^2}{\Delta^2} \right).
\end{equation}
The first term on the right comes from the outer-mode bound state and is independent of the Zeeman energy $B$. The second term originates from the inner-mode bound state. At the topological transition ($B = \Delta$) it vanishes along with the inner-mode bound state, and
\begin{equation}
\label{eq:disc_trans}
    \delta I_{B = \Delta} = e\Delta;
\end{equation}
see Fig.~\ref{fig:josephson}(b).

In the topological regime, $B>\Delta$, the zero-temperature Josephson current can be calculated from Eq.~\eqref{eq:JC} by using Eqs.~\eqref{eq:EgsEgs1Egs2}, \eqref{eq:cont}, and \eqref{eq:Egs_top_1}.
A representative current-phase relation for $B > \Delta$ is shown in Fig.~\ref{fig:josephson}(c).
In the topological state, the discontinuities in the Josephson current occur at $\phi=\pm(\pi+\phi_0)$ following the abrupt change in the ground state structure.
The displacement of the steps in $I(\phi)$ from $\phi = \pm \pi$ is a manifestation of the $4\pi$-periodic Josephson effect. The magnitude of the discontinuities is given by
\begin{equation}
\label{eq:disc_top}
    \delta I_{B>\Delta} = e\Delta\sqrt{1-(1-B/\Delta)^2} .
\end{equation}
Close to the topological transition, $B -\Delta \ll \Delta$,
\begin{subequations}
\begin{align}
    &\phi_0 \approx 2(B-\Delta)/\Delta\ll 1, \\* 
    &\delta I_{B>\Delta} \approx e\Delta .
\end{align}
\end{subequations}
Therefore, for $|\Delta - B| \ll \Delta$ the positions of the steps and their magnitude differ only by small amounts on the two sides of the transition. Consequently, in spite of the dramatic change in the ground state wave-function, the Josephson current $I(\phi)$ changes gradually across the topological phase transition. More generally, the $4\pi$-periodic Fourier harmonics in thermodynamic quantities build up continuously at $B>\Delta$, departing from zero at $B=\Delta$.

To conclude this section, we highlight an additional interesting property of the Josephson effect in the presence of Zeeman splitting. On both sides of the transition, there exists a nonzero, $2\pi$-periodic, contribution $I^{(2)}(\phi) = 2e\,dE_\mathrm{gs}^{(2)}\!/d\phi$ that comes from the continuum states.
For $B\sim\Delta$ this contribution is of the same order as the one coming from the Andreev bound states.
We note, however, that $I^{(2)}(\phi)$ is smooth and thus has no effect on the discontinuities in the current (see pale dashed lines in Fig.~\ref{fig:josephson}).
Close to the transition point, $|\Delta - B| \ll \Delta$, $I^{(2)}(\phi)$ can be computed analytically by differentiating Eq.~\eqref{eq:cont_vic} with respect to $\phi$.
This implies that $I(\phi)$ has a logarithmic contribution $\propto (\Delta-B)\ln(\Delta/|\Delta - B|)$.

\section{Effects of scattering at the junction}
\label{sec:scat}

In this section we discuss the influence of scattering on spectral (Sec.~\ref{sec:states_scat}) and thermodynamic (Sec.~\ref{sec:thermo_scat}) properties of the junction, focusing primarily on the case of weak backscattering. 
The backscattering couples inner and outer modes, modifying the structure of the Andreev states below the gap (Secs.~\ref{sec:bs_scatt_triv}, \ref{sec:bs_scatt_top}) as well as of the states above the continuum edge (Sec.~\ref{sec:cont_scatt}).
However, unlike the normal state (zero-field) conductance, which is solely determined by the backscattering strength, $1 - D$, the spectrum of the junction is also sensitive to the forward scattering phase $\gamma$.
Such sensitivity is especially prominent in the topological regime.
There, even for weak backscattering, a large forward scattering phase can have a dramatic qualitative impact on the Andreev levels (see Fig.~\ref{fig:andreev_spectrum_fwd_scatt}) and the Josephson current (see Fig.~\ref{fig:current_bs}). Motivated by this peculiarity, we also consider analytically the energy spectrum at $D = 1$, $\gamma \neq 0$ in Appendix~\ref{app:states_fwsc}.

\subsection{Bound states and continuum states in the presence of scattering}
\label{sec:states_scat}

The boundary condition \eqref{eq:BdG_bc} indicates that, in contrast to the case of the transparent junction, the inner and outer modes cannot be considered separately at $D < 1$.
In such a setting it is convenient to further employ the scattering approach to describe the spectrum of the system. A solution to the scattering problem yields scattering amplitudes which we arrange in the $S$-matrix, $S(E,\phi)$. In virtue of unitarity, above the gap, $E > |\Delta - B|$, the determinant of the 
scattering matrix  can be parameterized as
\begin{equation}
\label{eq:detS_lambda}
    \det S(E,\phi) = \frac{\Lambda^\star(E,\phi)}{\Lambda(E,\phi)}.
\end{equation}
The structure of the function $\Lambda(E,\phi)$ in the complex energy plane contains full information about the spectrum of the Josephson junction. The branch cuts of $\Lambda(E,\phi)$ situated at the real axis correspond to the states of the quasiparticle continuum with $|E| > |\Delta - B|$. In this range, the scattering induced correction to the continuum density of states $\delta \rho (E,\phi)$ can be found from $\Lambda(E,\phi)$ through Eqs.~\eqref{eq:rho_S} and \eqref{eq:detS_lambda}.
Zeros of $\Lambda(E,\phi)$ on the real axis within the interval $|E| < |\Delta - B|$ correspond to the energies of bound states, i.e., the latter can be found from
\begin{equation}
\label{eq:bs_equation}
    \Lambda(E,\phi) = 0,\qquad |E| < |\Delta - B|.
\end{equation}

\begin{figure*}[t]
  \begin{center}
    \includegraphics[width=0.97\textwidth]{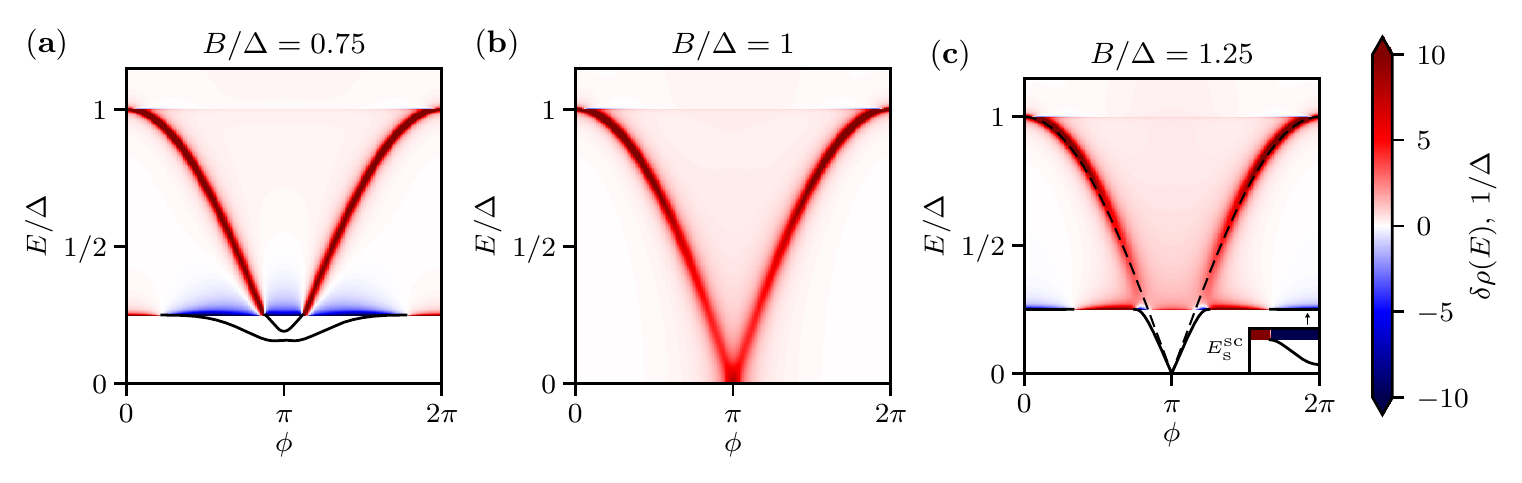}
    \caption{Evolution of the single-particle excitation spectrum of the junction with increasing Zeeman energy, in the presence of weak scattering.
    This is the analog of Fig.~\ref{fig:andreev_levels} for nonzero backscattering. 
    The scattering parameters used are $D = 0.925$, $\gamma \approx - 0.09 \pi$ (the relation between the parameters corresponds to the case of a delta function barrier, see Eq.~\eqref{eq:deltabarrier}).
    Bound states are depicted with solid black curves below the gap. The dashed curve in panel (c) shows the unperturbed bound state energy $\pm E_0(\phi)$. The inset in panel (c) is a close-up look at the vicinity of the continuum edge near $\phi = 2\pi$; there, a shallow bound state with energy $E_\mathrm{s}^\mathrm{sc}(\phi)$ is present below the continuum edge (the $\phi$-axis of the inset is to scale, the energy window is $2 \cdot 10^{-3}\Delta$). A similar state is present symmetrically close to $\phi = 0$. Above the gap, $E > |\Delta - B|$, the color corresponds to the scattering-induced correction to the density of states, $\delta \rho(E,\phi)$.}
    \label{fig:andreev_spectrum_backscattering}
  \end{center}
\end{figure*}

To determine $\Lambda(E,\phi)$ we solve the BdG equations \eqref{eq:BdG_equations} with boundary condition \eqref{eq:BdG_bc} and calculate the $S$-matrix.
The resulting expression for $\Lambda(E,\phi)$, valid at arbitrary backscattering strength $1-D$ and for any forward scattering phase $\gamma$, is explicitly presented in Appendix~\ref{app:full_spectrum} [see Eq.~\eqref{eq:app_full_Lambda}]. With its help the energies of the bound states and $\delta\rho(E,\phi)$ can be determined numerically at any value of $B/\Delta$ using Eqs.~\eqref{eq:rho_S}, \eqref{eq:detS_lambda}, and \eqref{eq:bs_equation}.
An example of such a numerical solution for moderately small $1 - D$ and $\gamma$ is presented in Fig.~\ref{fig:andreev_spectrum_backscattering}. The figure reveals a set of interesting features of the spectrum introduced by scattering at the junction, which we discuss below.
First, we concentrate on subgap energies, $|E| < |\Delta - B|$. We study the behavior of the Andreev levels on the trivial and topological sides of the transition in Secs.~\ref{sec:bs_scatt_triv} and \ref{sec:bs_scatt_top}, respectively. Then, in Sec.~\ref{sec:cont_scatt} we consider the effects of scattering on the spectrum above the continuum edge, $|E| > |\Delta - B|$.

\subsubsection{Bound states in the presence of scattering in the trivial phase, \texorpdfstring{$B < \Delta$}{B < Δ}}
\label{sec:bs_scatt_triv}

In the topologically trivial phase, $B < \Delta$, the inner and outer-mode bound states are hybridized by the scattering.
The degenerate zero-energy crossing which was present for the transparent junction at $\phi = \pi$ splits and gets pushed away from $E = 0$ at $D < 1$, $\gamma \neq 0$  [cf.~Fig.~\ref{fig:andreev_spectrum_backscattering}(a) and Fig.~\ref{fig:andreev_levels}(a)].

The hybridization of the bound states near $\phi = \pi$ can be addressed quantitatively in the limit
\begin{equation}
\label{eq:weak_sc_1}
    1-D \ll \frac{\abs{\Delta-B}}{\Delta} \ll 1, \quad \abs{\gamma} \ll 1,
\end{equation}
i.e., close to the topological transition threshold and for perturbatively weak scattering. In this limit, the general expression for $\Lambda(E,\phi)$ [given by Eq.~\eqref{eq:app_full_Lambda}] can be substantially simplified at $|E| \ll \Delta - B$ and $|\phi - \pi|\ll 1$ [see Eq.~\eqref{eq:app_lowest}]. Then, within the leading-order approximation, Eq.~\eqref{eq:bs_equation} yields bound states with energies $E = \pm E^\mathrm{sc}_{0,1}(\phi)$ where
\begin{align}
\label{eq:bs_hybr}
    E^\mathrm{sc}_{0,1}(\phi) 
    \approx \, &\pm \frac{E_1(\phi) - E_0(\phi)}{2} \nonumber \\* 
    &+ \frac{1}{2}\sqrt{[E_1(\phi)+E_0(\phi)]^2 + 8\Delta(\Delta - B)(1-D)}.
\end{align}
Here $E_{0}(\phi)$ and $E_{1}(\phi)$ are given by Eqs.~\eqref{eq:E0} and \eqref{eq:bs_energy}, respectively.
Eq.~\eqref{eq:bs_hybr} indicates that at $\phi = \pi$ the backscattering pushes the bound state energies away from zero by an amount
\begin{equation}\label{eq:deltaep}
    \delta \varepsilon\approx \sqrt{1-D}\,\sqrt{2\Delta(\Delta - B)}.
\end{equation}
Notice that $E^\mathrm{sc}_{0,1}(\phi)$ reaches the minimum $\sim (\Delta - B)\sqrt{1-D}$  at points symmetrically shifted away from $\phi=\pi$ by $\sim\sqrt{1 - D}$ [see Fig.~\ref{fig:andreev_spectrum_backscattering}(a)].
This shift of the minimum is known to occur for a junction with strong spin-orbit coupling even at zero magnetic field, but only away from the short junction limit \cite{chtchelkatchev2003,beri2008,tosi2019}.

We note that within the accuracy of Eq.~\eqref{eq:bs_hybr} the levels $E_{0,1}^\mathrm{sc}(\phi)$  cross at $\pi$. This is a peculiarity of the lowest-order perturbative calculation. In a subleading order, $|\gamma|\ll 1$ results in an anticrossing between $E^\mathrm{sc}_0(\phi)$ and $E^\mathrm{sc}_1(\phi)$ near $\phi = \pi$
with a gap $\sim |\gamma|\,\delta\varepsilon$ [see Fig.~\ref{fig:andreev_spectrum_backscattering}(a)]. 
Taking this anticrossing into account we find the bound state energies
\begin{align}
\label{eq:anticross}
    E_\mathrm{\pm}^\mathrm{sc}(\phi) \approx \ &\frac{E^\mathrm{sc}_0(\phi) + E^\mathrm{sc}_1(\phi)}{2} \nonumber \\*
    &\pm \frac{1}{2}\sqrt{[E^\mathrm{sc}_0(\phi) - E^\mathrm{sc}_1(\phi)]^2 + \gamma^2 \delta\varepsilon^2}.
\end{align}

\subsubsection{Bound states in the presence of scattering in the topological phase, \texorpdfstring{$B > \Delta$}{B > Δ}}
\label{sec:bs_scatt_top}

On the topological side of the transition, $B > \Delta$, the energy of the outer-mode bound state crosses zero at $\phi = \pi$ despite the presence of scattering [see Fig.~\ref{fig:andreev_spectrum_backscattering}(c)].
The robustness of the crossing is a consequence of fermion parity conservation \cite{fu2009}.
A strong modification of the bound state energy arises only when it approaches the edge of the continuum, driven by level repulsion between the bound state and states of the spectral continuum.
As we will show in Sec.~\ref{sec:thermo_scat}, this modification is important for thermodynamic properties of the junction.

To describe the repulsion of the outer-mode Andreev state from the continuum analytically, we again concentrate on the vicinity of the topological transition and assume that the scattering is perturbatively weak [Eq.~\eqref{eq:weak_sc_1}].
Then, the energy of the bound state $E_0^\mathrm{sc}(\phi)$ can be obtained by solving $\Lambda (E,\phi) = 0$ approximately in the limit $B - \Delta - |E| \ll B - \Delta$ (see Appendix~\ref{sec:app_top_bs}).
We find that the Andreev state merges with the continuum, i.e., its energy $E_0^\mathrm{sc}(\phi)$ reaches $|E|= B - \Delta$, at $\phi= \pi \pm \phi_0^\mathrm{sc}$ with
\begin{equation}
\label{eq:phi0sc}
    \phi_0^\mathrm{sc} \approx \phi_0 + 2(1-D),
\end{equation} 
and $\phi_0$ defined in Eq.~(\ref{eq:phi0}).
For small deviations from this point, $0 < (\pi + \phi_0^\mathrm{sc}) - \phi \ll \phi_0$, we obtain
\begin{multline}
\label{eq:boundsc}
    E^\mathrm{sc}_0(\phi) \approx -(B-\Delta) 
    \bigg\{ 1 -\frac{1}{2} \frac{\Delta^2}{(B-\Delta)^2} \\*
    \times \Bigl[\sqrt{(1-D)^2 + \tfrac{B-\Delta}{\Delta} (\pi + \phi_0^\mathrm{sc} - \phi)} -(1-D)\Bigr]^2 \bigg\} .
\end{multline}
Expression \eqref{eq:boundsc} verifies that the bound state energy is almost unperturbed far from the continuum: $E_0^\mathrm{sc}(\phi) \approx E_0(\phi)$ at $(\pi + \phi_0^\mathrm{sc}) - \phi \gg (1-D)^2 \Delta/(B-\Delta)$.
Conversely, at $(\pi + \phi_0^\mathrm{sc}) - \phi \ll (1-D)^2 \Delta/(B-\Delta)$ the level hybridization with the continuum is effective, and $E_0^\mathrm{sc}(\phi)$ reaches the continuum edge at $\phi = \pi + \phi_0^\mathrm{sc}$ with a zero slope.
Close to $\phi = \pi - \phi_0^\mathrm{sc}$ the behavior of $E_0^\mathrm{sc}(\phi)$ can be obtained from Eq.~\eqref{eq:boundsc} via  $E_0^\mathrm{sc}(\phi) = -E_0^\mathrm{sc}(2\pi -\phi)$ [see Fig.~\ref{fig:andreev_spectrum_backscattering}(c)].
Finally, we note that a small forward scattering phase $|\gamma|\ll 1$ results in a slight modification of the numeric prefactors in front of $1-D$ in Eqs.~\eqref{eq:phi0sc}, \eqref{eq:boundsc} (see Appendix~\ref{sec:app_top_bs}). Such a modification is suppressed in these expressions.

Another interesting feature of the Andreev spectrum at $B > \Delta$ is that weak scattering induces shallow bound states at
$\phi = 0$ and $2\pi$ [see inset in Fig.~\ref{fig:andreev_spectrum_backscattering}(c)]. These states merge with the edge of the continuum at a finite deviation $\phi_\mathrm{s}$ of phase $\phi$ from $0, 2\pi$. 

To quantify how these states peel off from the continuum, we compute $\Lambda(E,\phi)$ for $B-\Delta - |E| \ll B - \Delta$ in the limits $\phi \ll 1$ and $2\pi - \phi \ll 1$ (see Appendix~\ref{app:sbs}) under the assumption of weak scattering, $1 - D \ll 1$, $|\gamma|\ll 1$.
Then, through Eq.~\eqref{eq:bs_equation} we show that the shallow bound states are present for $\phi \in [0,\phi_\mathrm{s})$ and $\phi \in (2\pi - \phi_\mathrm{s}, 2\pi]$, where 
\begin{equation}
\label{eq:shallow_r}
    \phi_\mathrm{s} \approx 2\sqrt{1-D + \gamma^2}.
\end{equation}
At $\phi \in [0,\phi_\mathrm{s})$, we find the energy of the shallow state $E = \pm E^\mathrm{sc}_\mathrm{s}(\phi)$ with
\begin{equation}
\label{eq:shallow}
    E^\mathrm{sc}_\mathrm{s}(\phi) \approx (B - \Delta) 
    \Bigl(1-\frac{1}{128} \bigl[\phi_\mathrm{s}^2 - \phi^2\bigr]^2\Bigr).
\end{equation}
Near $\phi=2\pi$ the energy of the shallow bound state satisfies $E=\pm E^\mathrm{sc}_\mathrm{s}(2\pi - \phi)$. 

The shallow bound states are remarkably sensitive to the forward scattering phase $\gamma$. As one can see from Eq.~(\ref{eq:shallow_r}), they appear in the energy spectrum even in the sole presence of forward scattering ($D=1$, $\gamma\neq 0$). This case, for arbitrary $\gamma$, is considered in Appendix~\ref{app:states_fwsc}. In addition to detailed calculations, we provide there a qualitative explanation of the shallow states' origin.

\begin{figure}[t]
  \begin{center}
    \includegraphics[width=0.35\textwidth]{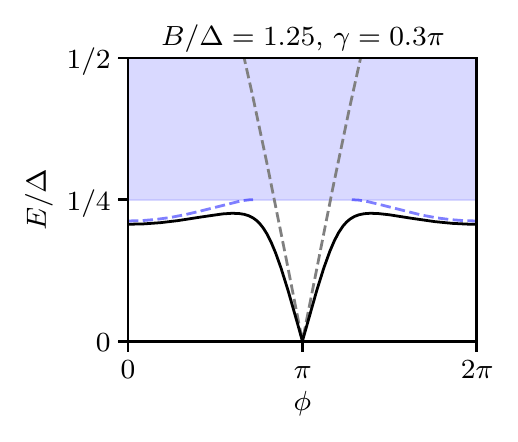}
    \caption{Single-particle excitation spectrum of the junction in the presence of a large forward scattering phase ($\gamma = 0.3\pi$) in the topological regime, $B = 1.25\Delta$.
    In the limit $D = 1$, two distinct shallow bound states are present near $\phi = 0,2\pi$ (dashed blue curves). They peel off from the continuum due to $\gamma \neq 0$. At $D < 1$ these states hybridize with the outer-mode bound state (dashed gray curve), resulting in an energy level that is separated from the continuum at all $\phi$ (solid black curve, $D = 0.925$). The spectrum is markedly different from that at $|\gamma|\ll 1$, cf.~Fig.~\ref{fig:andreev_spectrum_backscattering}(c).
    Note that the scattering-induced correction to the density of states above the gap is not shown here.}
    \label{fig:andreev_spectrum_fwd_scatt}
  \end{center}
\end{figure}

The domain of $\phi$ containing shallow states broadens with the increase of $|\gamma|$, as indicated by Eq.~\eqref{eq:shallow_r}. When $\gamma$ gets sufficiently close to $\pm\pi/2$, the shallow states hybridize with the outer-mode bound state due to weak backscattering (near the transition, $B - \Delta \ll \Delta$, this requires $|\gamma \pm \pi/2|\sim \sqrt{(B-\Delta)/\Delta}$, see Appendix~\ref{app:states_fwsc}).
As a result of the hybridization, an energy level is formed that is separated from the continuum at all phase differences, see Fig.~\ref{fig:andreev_spectrum_fwd_scatt}.
We note that such a strong dependence of the Andreev spectrum on the forward scattering phase is in sharp contrast to the normal state conductance of the junction; the latter is independent of the forward scattering phase and is determined only by the transmission probability $D$.

\subsubsection{Spectrum at \texorpdfstring{$|E| > |\Delta - B|$}{|E| > |Δ - B|} in the presence of scattering}
\label{sec:cont_scatt}

Next, we discuss the influence of scattering on the structure of the spectrum above the continuum edge.

In Sec.~\ref{sec:bs} we observed that, in the transparent junction limit, the Andreev state in the outer modes coexists with the inner-mode continuum at $|E| > |\Delta - B|$, see Fig.~\ref{fig:andreev_levels}(a)-(c). This coexistence is disrupted by weak backscattering: at $D < 1$ the outer-mode state hybridizes with the continuum states and broadens into a narrow resonance. The broadening is seen in Fig.~\ref{fig:andreev_spectrum_backscattering} as a peak in the density of states $\delta \rho(E,\phi)$ within the energy interval $E\in (|\Delta - B|,\Delta)$.

Precisely at the transition, $B = \Delta$, the bound state is broadened at all phase differences [see Fig.~\ref{fig:andreev_spectrum_backscattering}(b)]. In this case, the broadening can be concisely addressed analytically. The shape of the associated resonance in the density of states can be determined by computing $\Lambda(E,\phi)$ via Eqs.~\eqref{eq:rho_S} and \eqref{eq:detS_lambda}. Assuming that $1-D \ll 1$, $|\gamma|\ll 1$, $0< E\ll \Delta$, and $|\phi -\pi|\ll 1$, we find
\begin{equation}
\label{eq:Lambda}
    \Lambda(E,\phi) = 2iE\left([E + i \Delta (1-D)]^2 - E_0^2(\phi)\right).
\end{equation}
Then, for the corresponding contribution to the density of states we get
\begin{equation}
\label{eq:reso}
    \delta \rho(E,\phi) = \frac{1}{\pi} \sum_{s = \pm 1} \frac{\Gamma}{(E+sE_0(\phi))^2+\Gamma^2},
\end{equation}
where $\Gamma \approx (1-D) \Delta$.
This implies that $\delta\rho$ is a superposition of two Lorentzian peaks with width $\sim \Delta (1-D)$ centered around $E = \pm E_0(\phi)$.
In the subleading order, a small forward scattering phase, $|\gamma|\ll 1$, changes the prefactor in the expression for the width of the peaks, $\Gamma \approx \Delta (1 - D)(1 + \gamma^2)$.
However, it does not alter $\delta \rho$ qualitatively: the broadening of the bound state into a resonance relies on a coupling between inner and outer modes, which is provided by $D < 1$.
We note that expressions \eqref{eq:Lambda}, \eqref{eq:reso} are equally applicable away from the transition at $|\Delta - B| \ll \Delta$, for energies $|\Delta - B| \ll E \ll \Delta$.

\subsection{Thermodynamic properties in the presence of scattering}
\label{sec:thermo_scat}

In this section we discuss how scattering affects the thermodynamic properties of the junction.
We focus on the ground state energy  $E_\mathrm{gs}(\phi)$ (Sec.~\ref{sec:GS_backscattering}) and Josephson current $I(\phi)$ (Sec.~\ref{sec:current_scatt}) in the even fermion parity sector on the two sides of the topological transition.

\subsubsection{Ground state energy and many-body spectrum} \label{sec:GS_backscattering}

\begin{figure}[t]
  \begin{center}
    \includegraphics[width=0.5\textwidth]{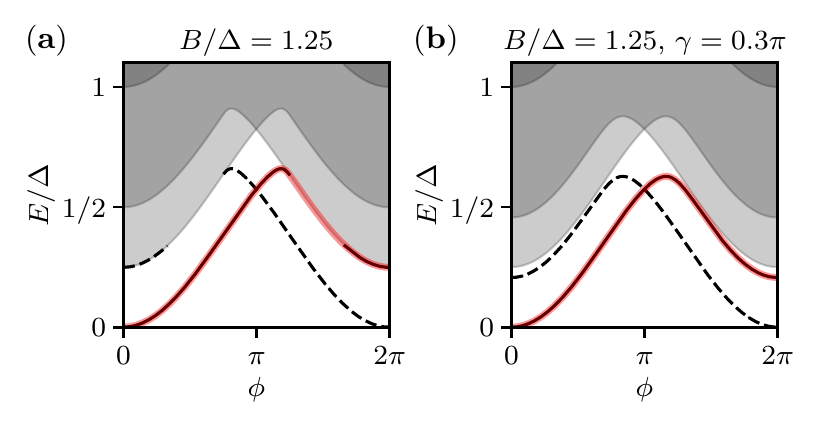}
    \caption{The influence of weak scattering on the ground state energy and many-body spectrum in the topological regime, $B = 1.25 \Delta$. Solid (dashed) curves represent many-body states with even (odd) fermion parity belonging to the discrete spectrum. Shaded grey regions represent the continuous part of the spectrum, of either fermion parity. The solid red curves denote the ground state energy in the even parity sector.
    At weak scattering [$D = 0.925$, $\gamma \approx - 0.09\pi$, panel (a)] there exists a phase region $\phi \in (\pi+\phi_0^\mathrm{sc}, 2\pi - \phi_\mathrm{s})$ where the energy gap between the ground state and the excited states is absent. For a larger forward scattering phase [$D = 0.925$, $\gamma = 0.3\pi$, panel (b)] the gap is present at all phases $\phi$.}
    \label{fig:many-body_spectrum_scattering}
  \end{center}
\end{figure}

We begin by studying the influence of scattering on $E_\mathrm{gs}(\phi)$ and on the many-body excitation energies of the junction. 
In the trivial regime, $B < \Delta$, 
the ground state at $D < 1$ is separated from the excitations by a gap at all phase differences.
The effects of the scattering on the many-body spectrum are more intricate in the topological regime, $B > \Delta$.
There, weak scattering {\it does not} destroy the phase domain (previously identified at $D=1$, $\gamma = 0$)  where the gap between the ground state and the excited  states is absent, see Fig.~\ref{fig:many-body_spectrum_scattering}(a). However,  this domain shrinks in the presence of scattering. On the one hand, close to $\phi = \pm 2\pi$ the shallow bound states peel off from the continuum and a small gap $\propto (B-\Delta)(1-D + \gamma^2)^2$ opens within $|\phi| \in (2\pi - \phi_\mathrm{s}, 2\pi]$.
On the other hand, backscattering shifts the points at which $E_\mathrm{gs}$ reaches the quasiparticle continuum from $\phi = \pm(\pi + \phi_0)$ to $\phi = \pm(\pi + \phi_\mathrm{0}^\mathrm{sc})$, where $\phi_\mathrm{0}^\mathrm{sc} > \phi_0$ [see Eq.~\eqref{eq:phi0sc}].
Overall, due to scattering the gapless domain spans the phase interval $|\phi|\in \left(\pi + \phi_0^\mathrm{sc}, 2\pi - \phi_\mathrm{s}\right)$, instead of $|\phi| \in (\pi + \phi_0, 2\pi]$ at $D = 1$, $\gamma = 0$.

This gapless domain might vanish completely if the scattering is sufficiently strong. In particular, at large forward scattering phase close to $\pm\pi/2$ the shallow bound states strongly hybridize with the outer-mode bound state, as discussed in Sec.~\ref{sec:bs_scatt_top}. As a result, the excitation spectrum is gapped at all phases [see Fig.~\ref{fig:many-body_spectrum_scattering}(b)].

\begin{figure*}[t]
  \begin{center}
    \includegraphics[width=0.85\textwidth]{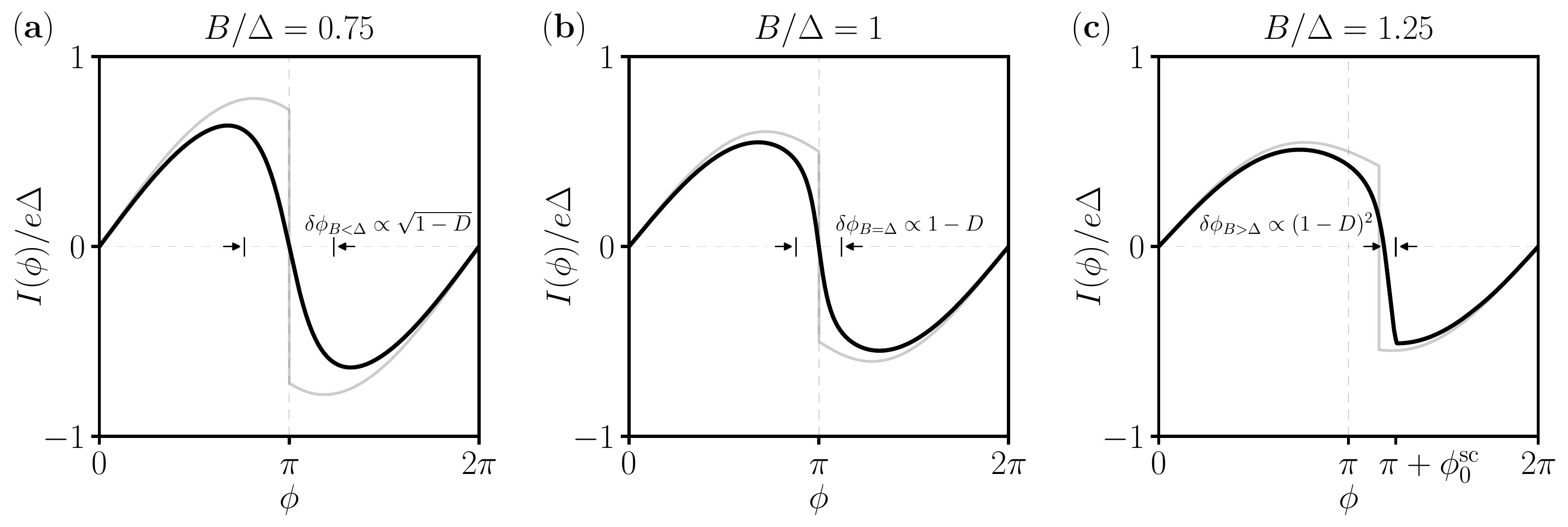}
    \caption{The influence of weak scattering on the Josephson current $I(\phi)$. Solid black curves depict $I(\phi)$ at $D = 0.925$, $\gamma \approx - 0.09 \pi$ (the relation between the parameters corresponds to the case of a delta function barrier, see Eq.~\eqref{eq:deltabarrier}). Solid grey curves are provided for reference and correspond to $I(\phi)$ for the transparent junction ($D=1$, $\gamma=0$; cf.~Fig.~\ref{fig:josephson}). The discontinuities present in $I(\phi)$ for the transparent junction get smeared by $D < 1$ in accordance with Eqs.~\eqref{eq:smear_triv}, \eqref{eq:smear_top}, and \eqref{eq:smear_thresh}. In the trivial regime [$B < \Delta$, panel (a)] and at the transition threshold [$B = \Delta$, panel (b)] the jump smears symmetrically. In the topological regime [$B > \Delta$, panel (c)] the jump, displaced from $\phi = \pi$, smears asymmetrically and turns into a kink at $\phi = \pi + \phi_0^\mathrm{sc}$.}
    \label{fig:current_bs}
  \end{center}
\end{figure*}

\subsubsection{Josephson current}
\label{sec:current_scatt}

The Josephson current $I(\phi)$ can be obtained from $E_\mathrm{gs}(\phi)$ through Eq.~\eqref{eq:JC} (see Appendix \ref{app:jc}). Numerically computed examples of $I(\phi)$ at weak scattering, $1 - D$, $|\gamma|\ll 1$, are presented in Fig.~\ref{fig:current_bs}. The plots highlight that scattering smears the discontinuities in $I(\phi)$ which were previously revealed at $D = 1$, $\gamma = 0$ (pale curves in Fig.~\ref{fig:current_bs}). This smearing happens differently on the trivial and topological sides of the transition.

In the trivial phase, $B < \Delta$, the discontinuity at $\phi =  \pi$ smears symmetrically in $\phi$ [see Fig.~\ref{fig:current_bs}(a)].
The smearing can be captured analytically in the vicinity of the topological transition, for perturbatively weak scattering [Eq.~\eqref{eq:weak_sc_1}].
By using Eq.~\eqref{eq:anticross} for $|\phi -\pi|\ll 1$ we find
\begin{equation}
\label{eq:trivialIsm}
    I(\phi) \approx \frac{e\Delta}{2} \frac{ (\pi - \phi)}{\sqrt{(\pi - \phi)^2 + 32(\Delta - B) (1-D)/\Delta}}.
\end{equation}
Here we neglected a small contribution of the continuum states to $I(\phi)$, as it weakly alters the result close to $\phi=\pi$ (see Appendix \ref{app:I2est} for details). 
Eq.~\eqref{eq:trivialIsm} indicates that the Josephson current interpolates  between $+e\Delta/2$ and $-e\Delta/2$ gradually over the phase interval
\begin{equation}
\label{eq:smear_triv}
    \delta\phi_{B<\Delta} \sim \left(\frac{\Delta - B}{\Delta}\right)^{\!\!1/2} \! (1-D)^{1/2}.
\end{equation}
The scaling of the smearing, $\delta\phi \propto \sqrt{1-D}$, is similar to the case of a regular short SNS junction \cite{beenakker1991}. We remark that the smearing relies on the presence of backscattering, whereas $|\gamma|\ll 1$ merely gives a subleading correction to the numeric prefactors in front of $1- D$ in Eqs.~\eqref{eq:trivialIsm}, \eqref{eq:smear_triv}; such corrections are omitted there.

Next, we discuss how scattering smears the discontinuity in $I(\phi)$ in the regime of the $4\pi$-periodic Josephson effect, $B>\Delta$ [Fig.~\ref{fig:current_bs}(c)].
The smearing can again be captured analytically in the vicinity of the transition for perturbatively weak scattering [Eq.~\eqref{eq:weak_sc_1}].
By using Eq.~\eqref{eq:boundsc} for $|\pi + \phi_0^\mathrm{sc} - \phi| \ll 1$ we find (see Appendix~\ref{app:jc_aprx_top} for details)
\begin{align}
\label{eq:I_top}
    I(\phi) \approx \ &e\Delta \Bigg[1 - \frac{1}{\sqrt{1 + \frac{B-\Delta}{\Delta(1-D)^2}(\pi + \phi_0^\mathrm{sc}-\phi)}}\Bigg] \nonumber \\*
    &\times \Theta(\pi + \phi_0^\mathrm{sc} - \phi) - \frac{e\Delta}{2}.
\end{align}
Backscattering replaces the discontinuity in the Josephson current at $\phi=\pi+\phi_0$ by a kink at $\phi=\pi+\phi_0^{\rm sc}$. The resulting function $I(\phi)$ does not have a symmetry around $\phi=\pi+\phi_0$. The overall smearing of the discontinuity occurs within the interval
\begin{equation}
\label{eq:smear_top}
    \delta\phi_{B>\Delta} \sim \frac{\Delta}{B - \Delta} (1 - D)^2.
\end{equation}
Comparing Eq.~(\ref{eq:smear_top}) with Eq.~(\ref{eq:smear_triv}), we see that the step in the Josephson current becomes {\it sharper} upon the transition to the topological phase, see Fig.~\ref{fig:current_bs}.
This happens due to the difference in the underlying physical mechanisms of the smearing.
In the trivial state the jump is smeared due to coupling between the bound states, whereas in the topological state the smearing is due to hybridization between the bound state and the continuum.

\begin{figure}[t]
  \begin{center}
    \includegraphics[width=0.4\textwidth]{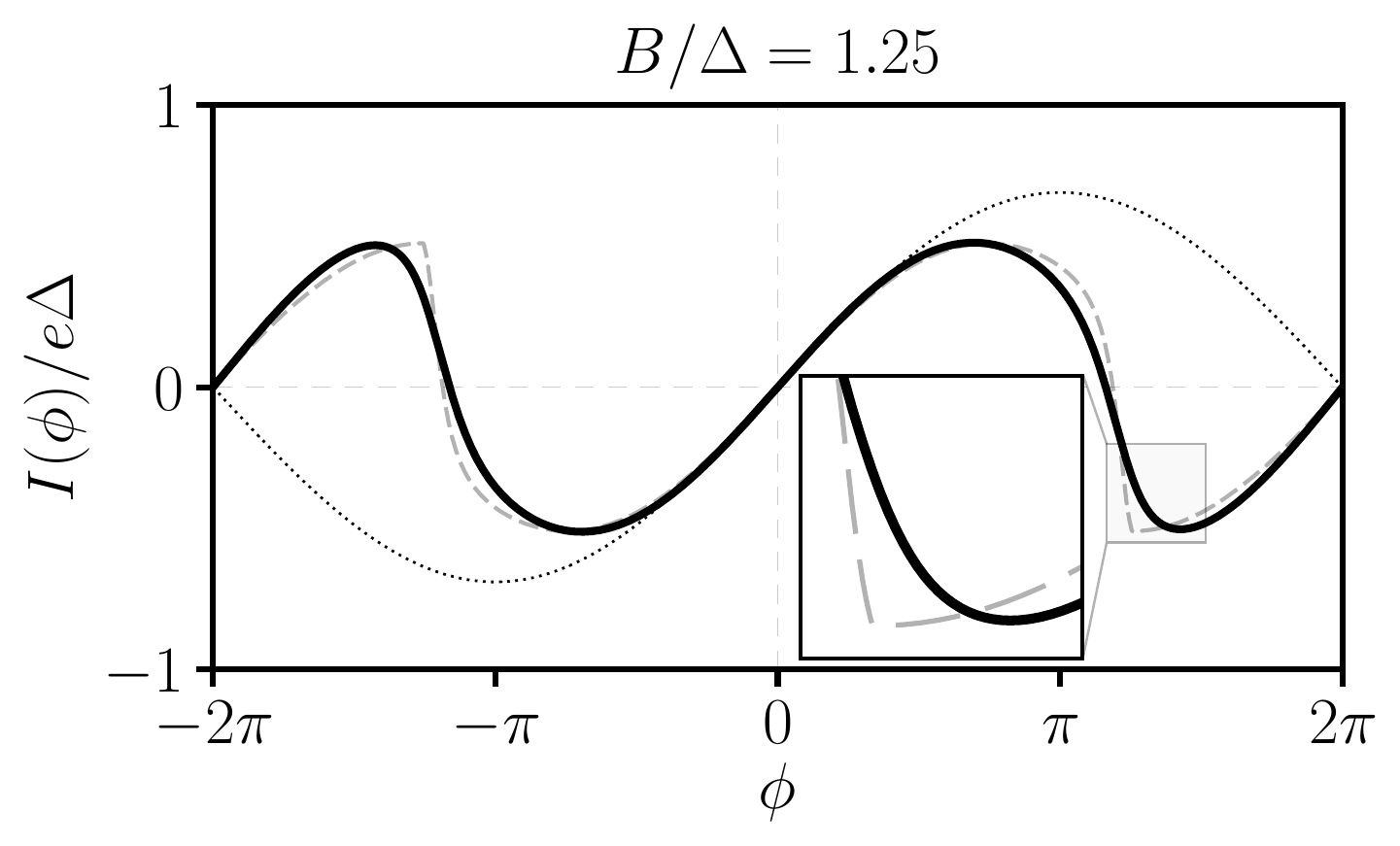}
    \caption{The influence of a large forward scattering phase $\gamma$ on the Josephson current $I(\phi)$ in the topological regime, $B = 1.25\Delta$. The solid black curve depicts $I(\phi)$ at $\gamma = 0.3\pi$ and $D = 0.925$. It is a smooth $4\pi$-periodic function of $\phi$; the kink which was present in $I(\phi)$ at smaller $\gamma$ [dashed gray curve, $\gamma \approx - 0.09\pi$ as in Fig.~\ref{fig:current_bs}(c)] is smeared out. The dotted curve corresponds to the dependence $I(\phi) \sim \sin(\phi /2)$ characteristic of the $4\pi$-periodic Josephson effect deep in the topological phase. 
    The inset is a close-up look at how the kink in $I(\phi)$ is smeared out by sufficiently large $\gamma$.}
    \label{fig:strong_bs}
  \end{center}
\end{figure}

At small backscattering, the kink in $I(\phi)$ persists for almost all values of $\gamma$. 
It may vanish only at $\gamma$ close to $\pm\pi/2$, when the shallow bound states merge with the outer-mode bound state and the many-body excitation spectrum becomes gapped at all phases [see Figs.~\ref{fig:andreev_spectrum_fwd_scatt} and \ref{fig:many-body_spectrum_scattering}(b)].
Then, the Josephson current turns into a smooth function of $\phi$ with $4\pi$-periodicity, see Fig.~\ref{fig:strong_bs}.

Finally, it is illuminating to consider how weak scattering influences the Josephson current at the topological transition threshold, $B = \Delta$.
At this point, there are no bound states in the system; only the broadened resonances are present. Thus, the behavior of the ground state energy and Josephson current is fully determined by the contribution from the continuum states.
In particular, the smearing of the discontinuity in $I(\phi)$ [see Fig.~\ref{fig:current_bs}(b)] can be studied with the help of Eq.~\eqref{eq:reso} in the weak scattering limit, $1-D \ll 1$, $|\gamma|\ll 1$.
For $|\phi - \pi|\ll 1$ we obtain the current step
\begin{equation}
    I(\phi) \approx \frac{e\Delta}{\pi} \arctan{\frac{\pi - \phi}{2(1-D)}} ,
\end{equation}
smeared on the scale
\begin{equation}
\label{eq:smear_thresh}
    \delta\phi_{B = \Delta} \sim 1 - D.
\end{equation}
This intermediate scale matches both with Eq.~\eqref{eq:smear_triv} and Eq.~\eqref{eq:smear_top} that describe the step width at the two sides of the transition, in its vicinity $|\Delta - B|/\Delta \sim 1 - D$.

\section{Summary and conclusions}
\label{sec:conclusions}

The Josephson effect is one of the most important probes of a superconductor since
it directly measures the order parameter.
It can also be used to distinguish topological superconductors from conventional ones, but the distinction is smeared by fluctuations of the junction's fermion parity caused by unpaired fermions in the bulk, such as those induced by disorder, 
quasiparticle poisoning, or finite length of the wire.
In this paper, we have shown that, even in the absence of
such fermion parity fluctuations, the behavior of a
highly-transparent Josephson junction is quite subtle
near the magnetic-field-driven topological transition in a proximitized semiconducting nanowire.
A qualitative distinction remains between the topological ($B > B_{\rm c}$, where $B_\mathrm{c} = \Delta$ in our model) and non-topological ($B < B_{\rm c}$) phases: the ground state energy and the Josephson current, $I(\phi)$, are $2\pi$-periodic in the latter case and $4\pi$-periodic
in the former.
However, the $4\pi$-periodicity of the Josephson current  just above the transition emerges as a shift of the zero-crossing in $I(\phi)$ away from $\phi = \pi$ by a small amount $\propto (B - B_{\rm c})/B_{\rm c}$. 
Consequently, the $4\pi$-periodic component of $I(\phi)$ develops gradually on the background of a larger $2\pi$-periodic component as the Zeeman energy $B$ is increased past $B_{\rm c}$.
For a multi-channel junction, the $4\pi$-periodic component of the Josephson
current will be accompanied by an even larger $2\pi$-periodic component due to the
non-topological bands in the wire, further masking the topological phase in the vicinity of the transition.

Nevertheless, there is an unusual and robust signature of the topological transition in a nanowire Josephson junction at high transmission. 
In the topological state, the excitation spectrum (within a fixed parity sector) is gapless over a range of phase differences, $|\phi|> \pi+\phi_0^\mathrm{sc}$.
This feature survives in the presence of weak scattering.
The gaplessness arises because, in that range of $\phi$, the bulk gap is smaller than the coupling between the two Majorana modes at the junction.
Consequently, the system lowers its energy by changing the fermion parity
of the coupled Majorana modes (relative to their parity at
$|\phi|< \pi+\phi_0^\mathrm{sc}$) and creating an above-gap fermionic excitation. 
An additional even number of channels cannot change this conclusion: it is possible to flip the parities of channels in pairs, but there will always be one channel left over that cannot change its parity to minimize the energy without creating an above-gap excitation. 

The gaplessness has observable implications for the Josephson current. It leads to kinks in the current-phase relation at the boundaries of the gapless domains, $\phi = \pm(\pi + \phi_0^\mathrm{sc})$. The presence of kinks in the topological phase is in contrast to the smooth $I(\phi)$ dependence in the non-topological regime. 

Finally, consistent with the notion of a phase transition, the ground state energy and critical current are non-analytic functions of $(B - B_\mathrm{c})$; we find a non-analytic contribution $\propto (B_\mathrm{c} - B) \ln \bigl(B_\mathrm{c}/|B_\mathrm{c} - B|\bigr)$ to $I(\phi)$.

Some signatures of the $4\pi$-periodic Josephson effect have so far been reported in dynamic properties of the junction, such as Shapiro steps \cite{rokhinson2012} or the microwave emission spectrum \cite{laroche2019}.
However, the results in this paper show that a large $2\pi$-periodic component is expected above but near the topological phase transition even for a single-channel junction.
This further complicates the interpretation of these experiments, which is already non-trivial due to quasiparticle poisoning.
We hope that our work will encourage further experimental activity aimed at detecting the critical point itself, in conjunction with the onset of Majorana signatures \cite{mourik2012, zhang2018, albrecht2015, deng2016}.
Motivated by recent and exciting progress in the microwave spectroscopy of Andreev bound states \cite{bretheau2013, vanwoerkom2016, hays2018, tosi2019, hays2019}, we will further investigate the signatures of the transition on the microwave absorption spectrum \cite{tewari2012,vayrynen2015} in an upcoming work.
Finally, the extension of the above findings to the case of finite voltage or current bias is another interesting direction for future research. \\[0.5em]

\acknowledgments

We acknowledge stimulating discussions with Angela Kou, Thorvald Larsen, Charlie Marcus, and Jay Sau.
CM and BvH thank Bela Bauer, William Cole and Anna Keselman for collaboration on related projects.
This work is supported by the DOE contract DE-FG02-08ER46482 (LIG), by the ARO grant W911NF-18-1-0212 (VK), and by the Microsoft corporation (CM).

\bibliography{references}

\clearpage
\appendix

\begin{widetext}
\section{Spectrum of inner modes at zero backscattering, \texorpdfstring{$D=1$}{D = 1}}
\label{app:imode_spectrum}

In this Appendix, we describe in detail the solution of the linearized BdG problem defined by Eqs.~\eqref{eq:BdG_equations} and \eqref{eq:BdG_bc}, for the case of zero backscattering, $D=1$. 
This leads to the results presented in Sec.~\ref{sec:transparent} of the main text. It also leads to the generalization of those results---to the case of purely forward scattering---described in Appendix~\ref{app:states_fwsc}.
When $D=1$ the inner and outer modes decouple from each other and can be analyzed independently for any value of $\gamma$, as follows from Eq.~\eqref{eq:BdG_bc}.
Thus, in this Appendix we will fix the transmission probability $D=1$ but leave the forward scattering phase $\gamma$ arbitrary.

The BdG problem for the inner modes is defined by Eqs.~\eqref{eq:BdG_equations_a} and \eqref{eq:BdG_bc_a} with $D=1$. It reduces to the problem for the outer modes [Eqs.~\eqref{eq:BdG_equations_b} and \eqref{eq:BdG_bc_b} with $D=1$] upon taking $B \to 0$ and conjugating with $\sigma_x$.
Therefore we restrict attention to the inner modes in the following.

It is convenient to remove the superconducting phase difference $\phi$ from the BdG equation \eqref{eq:BdG_equations_a} and have it appear in the boundary condition \eqref{eq:BdG_bc_a} instead.
This is accomplished by the unitary gauge transformation
\begin{equation}
\label{eq:app_gauge_transf}
    \Phi(x) \to e^{i (\phi/4) \sgn(x) \tau_z} \Phi(x) .
\end{equation}
The transformed BdG equation and boundary condition (at $D=1$) for the inner modes are then
\begin{equation}
\label{eq:app_imode}
    \big[ -i\alpha \pd_x \tau_z \sigma_z - B \sigma_x + \Delta \tau_x \big] \Phi(x) = E \Phi(x)
\end{equation}
and
\begin{equation}
\label{eq:app_imode_bc}
    \Phi(0^+) = e^{-i(\phi/2) \tau_z} e^{i\gamma \sigma_z} \Phi(0^-) .
\end{equation}
Since we deal exclusively with the inner modes in this Appendix, we omit the subscript ``$\mathrm{i}$'' on the wave function $\Phi$.
We solve Eqs.~\eqref{eq:app_imode} and \eqref{eq:app_imode_bc} by a standard wavefunction matching procedure, detailed below.

First consider the auxiliary translation-invariant problem
\begin{equation}
\label{eq:app_imode_Hkphi}
    \mc{H}(-i\pd_{x}) \, \Phi(x) = E \Phi(x) ,
\end{equation}
where
\begin{equation}
    \mc{H}(k) \equiv \alpha k \tau_z \sigma_z - B \sigma_x + \Delta \tau_x .
\end{equation}
For each fixed $E$, the space of vector-valued functions $\Phi(x)$ (in spin and Nambu space) that solve Eq.~\eqref{eq:app_imode_Hkphi} (with no boundary conditions imposed at spatial infinity) is four-dimensional; the solutions are generalized plane waves of the form $\Phi_n(E) e^{ik_n(E) x}$, with wavevectors $k_n(E)$ that are roots of the characteristic polynomial
\begin{equation}
    \det[E - \mc{H}(k)] = [E^2 - (\alpha k)^2 - (\Delta + B)^2] [E^2 - (\alpha k)^2 - (\Delta - B)^2] .
\end{equation}
The four roots are $\sigma k_{\nu}(E)$, where $\sigma,\nu = \pm 1$, and
\begin{equation}
\label{eq:app_imode_kE}
    k_{\nu}(E) = \frac{1}{\alpha} \, [E^2 - (\Delta + \nu B)^2]^{1/2} .
\end{equation}
These are complex in general; by convention, the square root is to be taken so that $k_{\nu}(E)$ always lies in the upper-right quadrant of the complex $k$-plane ($\re k \geq 0$, $\im k \geq 0$).
The associated vectors $\Phi_{\sigma,\nu}(E)$ solve
\begin{equation}
    [E - \mc{H}(\sigma k_{\nu}(E))] \Phi_{\sigma,\nu}(E) = 0 .
\end{equation}
They are given by
\begin{equation}
\label{eq:app_imode_Phi}
    \Phi_{\sigma,+}(E) = 
    \frac{1}{2\sqrt{i \sin{\eta_+(E)}}} \begin{bmatrix}
    e^{i\sigma \eta_+(E)/2} \\
    - e^{-i\sigma \eta_+(E)/2} \\
    e^{-i\sigma \eta_+(E)/2} \\
    - e^{i\sigma \eta_+(E)/2}
    \end{bmatrix} \! , \qquad
    \Phi_{\sigma,-}(E) = 
    \frac{1}{2\sqrt{i \sin{\eta_-(E)}}} \begin{bmatrix}
    e^{i\sigma \eta_-(E)/2} \\
    \sgn(\Delta - B) \, e^{-i\sigma \eta_-(E)/2} \\
    \sgn(\Delta - B) \, e^{-i\sigma \eta_-(E)/2} \\
    e^{i\sigma \eta_-(E)/2}
    \end{bmatrix} \! ,
\end{equation}
where
\begin{equation}
\label{eq:app_imode_etaE}
    e^{i\eta_{\nu}(E)} = \frac{E + \alpha k_{\nu}(E)}{\abs{\Delta + \nu B}} .
\end{equation}
We choose branches so that $\eta_{\nu}(E)$ is a continuous function of $E$ that approaches $-i \infty$ as $E \to +\infty$. Thus, $\eta_{\nu}(E)$ is purely imaginary for $E > \abs{\Delta + \nu B}$, and real for $\abs{E} < \abs{\Delta + \nu B}$.

The normalization factors $\frac{1}{2} (i \sin{\eta_{\nu}(E)})^{-1/2}$ in Eq.~\eqref{eq:app_imode_Phi} are fixed by requiring that the associated plane waves carry unit probability current $j_P \equiv \Phi^{\dagger} (\tau_z \sigma_z) \Phi$ at energies $E > \abs{\Delta + \nu B}$ (where these modes are propagating).
This is necessary to obtain a unitary scattering matrix.

Any function $\Phi(x;E)$ that satisfies Eq.~\eqref{eq:app_imode} can be expressed as a linear combination of the generalized plane waves at that energy $E$, as follows:
\begin{equation}
\label{eq:app_imode_exp}
    \Phi(x;E) = \begin{cases}
    a_{-,\nu} \, \Phi_{-,\nu}(E) e^{-ik_{\nu}(E) x} 
    + b_{+,\nu} \, \Phi_{+,\nu}(E) e^{+ik_{\nu}(E) x} , &\quad x > 0 , \\[0.1em]
    a_{+,\nu} \, \Phi_{+,\nu}(E) e^{+ik_{\nu}(E) x} 
    + b_{-,\nu} \, \Phi_{-,\nu}(E) e^{-ik_{\nu}(E) x} , &\quad x < 0 ,
    \end{cases}
\end{equation}
where a summation over $\nu = \pm$ is implicit, and $k_{\nu}(E)$, $\Phi_{\sigma,\nu}(E)$ are given in Eqs.~\eqref{eq:app_imode_kE}, \eqref{eq:app_imode_Phi} respectively.
When the wavevectors are real, the coefficients $a_{\sigma,\nu}$ ($b_{\sigma,\nu}$) multiply incoming (outgoing) plane waves.
When the wavevectors are imaginary, $a_{\sigma,\nu}$ ($b_{\sigma,\nu}$) multiply the generalized plane waves that grow (decay) exponentially at infinity.

The boundary condition at $x=0$, Eq.~\eqref{eq:app_imode_bc}, then yields a linear system of equations to be satisfied by the coefficients:
\begin{equation}
\label{eq:app_imode_bc_ab}
    \Omega_a(E,\phi,\gamma) \, \vec{a} = \Omega_b(E,\phi,\gamma) \, \vec{b} ,
\end{equation}
where
\begin{subequations}
\begin{align}
    \vec{a} &= (a_{++}, \, a_{+-}, \, a_{-+}, \, a_{--})^T , \\
    \vec{b} &= (b_{-+}, \ b_{--}, \ b_{++}, \ b_{+-})^T ,
\end{align}
\end{subequations}
and $\Omega_a, \Omega_b$ are $4 \times 4$ matrices, given by
\begin{subequations}
\begin{align}
    \Omega_a(E,\phi,\gamma) &= \big[ 
    e^{-i(\phi/2)\tau_z} e^{i\gamma \sigma_z} \Phi_{+,+}(E), \ 
    e^{-i(\phi/2)\tau_z} e^{i\gamma \sigma_z} \Phi_{+,-}(E), \ 
    - \Phi_{-,+}(E), \ 
    -\Phi_{-,-}(E) \big] , \\*
    \Omega_b(E,\phi,\gamma) &= \big[ 
    \! - \! e^{-i(\phi/2)\tau_z} e^{i\gamma \sigma_z} \Phi_{-,+}(E), \ 
    - e^{-i(\phi/2)\tau_z} e^{i\gamma \sigma_z} \Phi_{-,-}(E), \ 
    \Phi_{+,+}(E), \ 
    \Phi_{+,-}(E) \big] . 
    \label{eq:app_imode_OmegaB}
\end{align}
\end{subequations}
Here, $[\Phi, \Psi, \cdots]$ denotes the matrix whose first column is $\Phi$, second column $\Psi$, and so on.
The wavefunction $\Phi(x;E)$ in Eq.~\eqref{eq:app_imode_exp} represents a valid solution of the linearized inner-mode BdG problem if and only if (i) the coefficients $a_{\sigma,\nu}, b_{\sigma,\nu}$ satisfy Eq.~\eqref{eq:app_imode_bc_ab}, and (ii) $\Phi(x;E)$ satisfies appropriate conditions at $x = \pm \infty$.

\subsection{Bound states}
\label{app:imode_bound_state}

We first consider states belonging to the discrete spectrum (bound states). As usual, these correspond to \emph{normalizable} wavefunctions $\Phi(x)$. Normalizability requires that the wavevectors $\sigma k_{\nu}(E)$ appearing in the expansion \eqref{eq:app_imode_exp} all have positive (negative) imaginary part for $x > 0$ ($x < 0$).
It follows immediately that bound states cannot exist at energies $\abs{E} > \abs{\Delta + B}$, since then $k_{\nu}(E)$ is real according to Eq.~\eqref{eq:app_imode_kE}.

In the energy range $\abs{\Delta - B} < \abs{E} < \abs{\Delta + B}$, $k_-(E)$ is real and $k_+(E)$ is imaginary (with $\im k > 0$). A bound state with energy in this range would therefore have the form
\begin{equation}
    \Phi(x;E,\phi,\gamma) = \begin{cases}
    b_{+,+}(E,\phi,\gamma) \, \Phi_{+,+}(E) e^{+i k_+(E) x} , &\ x > 0 , \\
    b_{-,+}(E,\phi,\gamma) \, \Phi_{-,+}(E) e^{- ik_-(E) x} , &\ x < 0 .
    \end{cases}
\end{equation}
However, it is easy to verify, using Eq.~\eqref{eq:app_imode_Phi}, that this only satisfes Eq.~\eqref{eq:app_imode_bc} when $b_{+,+} = b_{-,+} = 0$ (in which case the wavefunction vanishes identically). 

Thus, bound states can exist only at energies $\abs{E} < \abs{\Delta - B}$. In this energy range, both $k_+(E)$ and $k_-(E)$ are imaginary (with $\im k > 0$), so the bound state wavefunction has the form \eqref{eq:app_imode_exp} with $a_{\sigma,\nu} \equiv 0$. The $b_{\sigma,\nu}$ coefficients must still satisfy Eq.~\eqref{eq:app_imode_bc_ab}, which now reduces to:
\begin{equation}
    0 = \Omega_b(E,\phi,\gamma) \, \vec{b} .
\end{equation}
This equation has a nontrivial solution for $\vec{b}$ (and hence a bound state exists at the given $E$ and $\phi$) if and only if $\det{\Omega_b(E,\phi,\gamma)} = 0$.
Explicit calculation using Eqs.~\eqref{eq:app_imode_Phi} and \eqref{eq:app_imode_OmegaB} yields
\begin{align}
\label{eq:app_imode_det_OmegaB}
    \det{\Omega_b(E,\phi,\gamma)} = 
    - \frac{1}{2\sin{\eta_+} \sin{\eta_-}} \Big( 
    &[\cos^2(\phi/2) + \cos^2\!\gamma] \sgn(\Delta - B) \sin{\eta_+} \sin{\eta_-} \nonumber \\*
    &- [\sin^2(\phi/2) + \sin^2\!\gamma] \sgn(\Delta - B) \cos{\eta_+} \cos{\eta_-} - [\sin^2(\phi/2) - \sin^2\!\gamma] \Big) .
\end{align}
We now define
\begin{equation}
\label{eq:app_imode_Lambda_def}
    \Lambda_{\mathrm{i}}(E,\phi,\gamma) = 
    - 2 \sin{\eta_+} \sin{\eta_-} 
    \frac{(\Delta + B)(\Delta - B)}{[\cos^2(\phi/2) + \cos^2\!\gamma]} \, \det{\Omega_b(E,\phi,\gamma)} .
\end{equation}
For $B \neq \Delta$, it is clear that $\det{\Omega_b(E,\phi,\gamma)} = 0$ if and only if $\Lambda_{\mathrm{i}}(E,\phi,\gamma) = 0$.
From Eq.~\eqref{eq:app_imode_etaE}, we have
\begin{subequations}
\begin{align}
    \abs{\Delta + \nu B} \cos{\eta_{\nu}(E)} &= E , \\
    \abs{\Delta + \nu B} \sin{\eta_{\nu}(E)} &= -i\alpha k_{\nu}(E) ,
\end{align}
\end{subequations}
Using these with Eq.~\eqref{eq:app_imode_det_OmegaB} in Eq.~\eqref{eq:app_imode_Lambda_def}, we obtain
\begin{equation}
\label{eq:app_imode_Lambda}
    \Lambda_{\mathrm{i}}(E,\phi,\gamma) = 
    - \alpha k_+(E) \, \alpha k_-(E) - F(\phi) [1 + G(\phi,\gamma)] (E^2 + \Delta^2 - B^2) - G(\phi,\gamma) \, (E^2 - \Delta^2 + B^2) .
\end{equation}
with
\begin{subequations}\label{eq:app_FG_def}
\begin{align}
    F(\phi) &\equiv
    \frac{\sin^2(\phi/2)}{\cos^2(\phi/2) + 1} , \\*
    G(\phi,\gamma) &\equiv
    \frac{\sin^2\!\gamma}{\cos^2(\phi/2) + \cos^2\!\gamma} .
\end{align}
\end{subequations}
Thus, the inner-mode bound state energy is a solution of the equation $\Lambda_{\mathrm{i}} = 0$, as stated in the main text.
Since $\alpha k_{\nu}(E) = i [(\Delta + \nu B)^2 - E^2]^{1/2}$ for $\abs{E} < \abs{\Delta + \nu B}$, Eq.~\eqref{eq:app_imode_Lambda} reduces to Eq.~\eqref{eq:Lambda_i} in the limit $\gamma \to 0$.
For general $\gamma \neq 0$, the bound state solutions of $\Lambda_{\mathrm{i}}(E,\phi,\gamma) = 0$ are analyzed in detail in Appendix~\ref{app:states_fwsc}.

\subsection{Scattering states}
\label{app:imode_s_matrix}

We next consider states belonging to the continuous spectrum (scattering states). 
For any values of $E, \phi, \gamma$ such that $\det \Omega_b(E,\phi,\gamma) \neq 0$, we may solve Eq.~\eqref{eq:app_imode_bc_ab} to express $\vec{b}$ as a linear function of $\vec{a}$:
\begin{equation}
    \vec{b} = [\Omega_b(E,\phi,\gamma)]^{-1} \, \Omega_a(E,\phi,\gamma) \, \vec{a} .
\end{equation}
The scattering states are then given by Eq.~\eqref{eq:app_imode_exp}. In order for the wavefunctions to remain finite as $x \to \pm \infty$, the coefficient $a_{\sigma,\nu}(E)$ must vanish unless the corresponding wavevector $k_{\nu}(E)$ is real. This condition fixes the number of linearly independent scattering states in the inner-mode continuum at energy $E$.

At energies $E > \Delta+B$, all waves are propagating (the wavevectors $k_{\nu}(E)$ are both real).
Thus, the inner-mode scattering matrix at these energies is the $4 \times 4$ matrix:
\begin{equation}\label{eq:app_Si_large_e}
    S_{\text{i}}(E,\phi,\gamma) = [\Omega_b(E,\phi,\gamma)]^{-1} \, \Omega_a(E,\phi,\gamma) .
\end{equation}
Explcit calculation shows that $\det \Omega_a(E,\phi,\gamma) = \det \Omega_b(E,\phi,\gamma)$. Therefore,
\begin{equation}
    \det S_{\text{i}}(E,\phi,\gamma) = 1, \quad\ E > \Delta + B.
\end{equation}

At energies in the range $\abs{\Delta - B} < E < \Delta + B$, only the ``$\nu = -$'' waves are propagating (the wavevector $k_-(E)$ is real, but $k_+(E)$ is imaginary).
Thus, the inner-mode scattering matrix at these energies is a $2 \times 2$ matrix relating $(b_{+-}, b_{--})^T$ to $(a_{+-}, a_{--})^T$.
It may be obtained as the appropriate sub-block of the $S$ matrix in Eq.~\eqref{eq:app_Si_large_e} (analytically continued below the threshold $E = \Delta + B$ by letting $\alpha k_+(E) \to i [(\Delta+B)^2 - E^2]^{1/2}$).
In this case, the determinant of $S_\mathrm{i}$ is given by
\begin{equation}
    \det S_{\text{i}}(E,\phi,\gamma) = 
    \frac{-\alpha k_-(E) \, \alpha k_+(E) + F(\phi) [1+G(\phi,\gamma)] \, (E^2 + \Delta^2 - B^2) + G(\phi,\gamma) \, (E^2 - \Delta^2 + B^2)}{+\alpha k_-(E) \, \alpha k_+(E) + F(\phi) [1+G(\phi,\gamma)] \, (E^2 + \Delta^2 - B^2) + G(\phi,\gamma) \, (E^2 - \Delta^2 + B^2)} ,
\end{equation}
where $F$ and $G$ are the functions defined in Eq.~\eqref{eq:app_FG_def}. Since $k_-(E)$ is real and $k_+(E)$ is imaginary, this can be rewritten as
\begin{equation}
    \det S_{\text{i}}(E,\phi,\gamma) = \frac{[\Lambda_{\mathrm{i}}(E,\phi,\gamma)]^\star}{\Lambda_{\mathrm{i}}(E,\phi,\gamma)} ,
\end{equation}
where $\Lambda_{\mathrm{i}}(E,\phi,\gamma)$ is given in Eq.~\eqref{eq:app_imode_Lambda}, and ${}^\star$ denotes complex conjugation. 
This reduces to Eq.~\eqref{eq:det_Si} in the limit $\gamma \to 0$.
For general $\gamma \neq 0$, the scattering-induced contribution to the continuum density of states, $\delta\rho(E,\phi,\gamma) = (2\pi i)^{-1} \partial_E \ln \det S_\mathrm{i}(E,\phi,\gamma)$, is analyzed in Appendix~\ref{app:states_fwsc}.

Finally, at energies $E < \abs{\Delta - B}$, there are no propagating waves (the wavevectors $k_{\nu}(E)$ are both imaginary), and so the inner-mode scattering matrix is not defined.

\clearpage
\end{widetext}

\section{Spectrum of the junction with purely forward scattering (\texorpdfstring{$D=1$}{D = 1} but \texorpdfstring{$\gamma \neq 0$}{ɣ ≠ 0})}
\label{app:states_fwsc}

In Secs.~\ref{sec:bs} and \ref{sec:cont} of the main text, we presented a detailed description of the Andreev spectrum in the nanowire junction setup in the absence of scattering ($D=1$ and $\gamma = 0$).
In this Appendix we generalize this discussion to the case of purely forward scattering ($D=1$ but $\gamma \neq 0$).
The generalization is natural on a technical level: when $D=1$, the inner and outer modes decouple from each other [as follows from Eq.~\eqref{eq:BdG_bc}], permitting a complete analytical solution of the BdG problem for any value of the forward scattering phase $\gamma$ (see Appendix~\ref{app:imode_spectrum} for details of the solution).
The results in this Appendix may be used to understand the Andreev spectra of short junctions that are characterized in the normal state by weak backscattering, $1-D \ll 1$, but large forward scattering phase, $\gamma \sim 1$, following the general discussion in Sec.~\ref{sec:scat} of the main text.

\subsection{The origin of the shallow states upon the topological transition}

As we already mentioned, the inner and outer modes decouple from each other at $D=1$.
The Andreev spectrum of the outer modes is independent of the forward scattering phase $\gamma$; this follows from the fact that $\gamma$ can be eliminated from the outer-mode problem [Eqs.~\eqref{eq:BdG_equations_b} and \eqref{eq:BdG_bc_b} with $D=1$] by a unitary transformation $\Phi_\mathrm{o}(x) \rightarrow e^{-i(\gamma/2) \sgn(x) \sigma_z} \, \Phi_\mathrm{o}(x)$. 
Thus, the contribution of the outer modes to the energy spectrum is not affected by $\gamma$.

The same unitary transformation applied to the inner modes absorbs the forward-scattering phase at the price of rotating the magnetic fields on the two sides of the junction in directions opposite to each other (the angle between the two directions is proportional to the forward-scattering phase shift). At some phase shift, the fields point in opposite directions. In the absence of superconductivity ($\Delta=0$), the corresponding Hamiltonian is identical to a Dirac Hamiltonian in 1D with a mass $m(x)$ which changes sign at $x=0$. That brings about a localized zero-energy state. At an arbitrary magnetic field rotation angle, the localized state moves away from zero energy, but does not vanish. Inclusion of a finite but small gap ($\Delta<B$) does not destroy this state.

\subsection{Inner-mode bound states with forward scattering in the trivial phase, \texorpdfstring{$B < \Delta$}{B < Δ}}

The Andreev spectrum of the inner modes depends sensitively on $\gamma$.
The inner-mode bound state energy is a solution of the equation $\Lambda_{\mathrm{i}}(E,\phi,\gamma) = 0$, with $\Lambda_\mathrm{i}$ given by Eq.~\eqref{eq:app_imode_Lambda} (in which $\alpha k_+ \alpha k_- = - \sqrt{(\Delta + B)^2 - E^2} \sqrt{(\Delta - B)^2 - E^2} < 0$ at energies $|E| < |\Delta - B|$).
In order for the equation $\Lambda_\mathrm{i}(E,\phi,\gamma) = 0$ to admit a solution, the second term of Eq.~\eqref{eq:app_imode_Lambda} must be negative.
Simple algebra shows that this condition is equivalent to
\begin{equation}
\label{eq:app_bs_cond_fwsc}
    (\Delta - B)[\Delta \sin^2(\phi/2) - B \sin^2\!\gamma] > 0 .
\end{equation}
In the trivial phase, $\Delta > B$, the inner-mode bound state exists for $\phi \in (\phi_1, 2\pi - \phi_1)$, where
\begin{equation}
\label{eq:app_phi1}
    \phi_1 = 2 \arcsin\!\big(\sqrt{B/\Delta} \, \abs{\sin{\gamma}}\big) .
\end{equation}
Its energy, $E = \pm E_1(\phi,\gamma)$, may be obtained analytically by solving $\Lambda_\mathrm{i}(E,\phi,\gamma) = 0$ (squaring this gives a quadratic equation in $E^2$, of which precisely one root satisfies $0 < E^2 < (\Delta-B)^2$ in the specified range of $\phi$).
The result is more complicated than Eq.~\eqref{eq:bs_energy} for $E_1(\phi)$, but shares many basic features with the latter.
Firstly, $E_1(\phi,\gamma) = E_0(\phi)$ at $B = 0$, so the Andreev spectrum remains two-fold degenerate in the absence of the magnetic field.
Secondly, $E_1(\phi,\gamma)$ still crosses zero at $\phi =\pi$, simultaneously with $E_0(\phi)$.
Thirdly, $\abs{E_1(\phi,\gamma)}$ remains bounded by $\Delta - B$.
However, now $E_1(\phi_1,\gamma) = \Delta - B$ and $E_1(2\pi - \phi_1,\gamma) = B - \Delta$, so that the bound state merges with the edge of the continuum at $\phi = \phi_1$ and $2\pi - \phi_1$, and disappears when $\phi \in [0, \phi_1]$ or $\phi \in [2\pi - \phi_1, 2\pi]$.

\subsection{Inner-mode bound states with forward scattering in the topological phase, \texorpdfstring{$B > \Delta$}{B > Δ}}

In the topological regime, $B > \Delta$, weak forward scattering causes shallow bound states to appear in the inner modes near $\phi = 0$ and $2\pi$.
As discussed above, the equation $\Lambda_\mathrm{i}(E,\phi,\gamma) = 0$ admits solutions for some $E$ in the range $\abs{E} < \abs{\Delta - B}$ only if Eq.~\eqref{eq:app_bs_cond_fwsc} is satisfied.
When $\gamma = 0$ this condition prohibits bound states in the inner modes at $B > \Delta$, but for $\gamma \neq 0$ it permits them to exist near $\phi = 0$ and $2\pi$.
For sufficiently small $\gamma$ [in particular for $\abs{\gamma} < \arcsin(\sqrt{\Delta/B})$, cf.~Eq.~\eqref{eq:app_phi1}], these states exist only for $\phi \in [0, \phi_1)$ and $\phi \in (2\pi-\phi_1, 2\pi]$, and have energies close to $B - \Delta$; they touch the edge of the inner-mode continuum at $\phi = \phi_1$, $2\pi - \phi_1$ and disappear when $\abs{\phi - \pi} < \pi - \phi_1$.
The shallow states have energies $E = \pm E^\mathrm{sc}_\mathrm{s}(\phi)$, for which a simple expression can be obtained in the limit $(B-\Delta) - \abs{E} \ll (B-\Delta)$ 
In this limit, Eq.~\eqref{eq:app_imode_Lambda} reduces to
\begin{align}
    \tfrac{1}{2} \Lambda_{\mathrm{i}}(E,\phi,\gamma) \approx 
    \, &\sqrt{2\Delta B (B-\Delta)(B-\Delta-\abs{E})} \nonumber \\*
    &- \Delta (B-\Delta) \, F(\phi) [1 + G(\phi,\gamma)] \nonumber \\*
    &- B(B-\Delta) \, G(\phi,\gamma) .
\end{align}
Then for the solution to $\Lambda_{\mathrm{i}}=0$, we obtain
\begin{multline}
\label{eq:E_sbs_fwsc}
    E^\mathrm{sc}_\mathrm{s}(\phi) \approx (B-\Delta) \bigg( 1 - \frac{B}{2\Delta} \\
    \times \left[ \frac{\sin^2\!\gamma - (\Delta/B) \sin^2(\phi/2)}{\cos^2\!\gamma +\cos^2(\phi/2)} \right]^2 \bigg)
\end{multline}
for $\phi \in [0, \phi_1)$ and $\phi \in (2\pi - \phi_1, 2\pi]$. 
This approximate expression is valid as long as $B - \Delta - E^\mathrm{sc}_\mathrm{s}(\phi) \ll B - \Delta$.
Note that, for $|\gamma|\ll 1$, $\phi \ll 1$ and $B-\Delta \ll \Delta$, Eq.~\eqref{eq:E_sbs_fwsc} reduces to Eq.~\eqref{eq:shallow} [with $\phi_{\mathrm{s}}$ given by Eq.~\eqref{eq:shallow_r} with $D=1$].
As $\abs{\gamma}$ increases, the shallow states persist over a larger range of $\phi$.
They merge into a single level when $\abs{\gamma} = \arcsin(\sqrt{\Delta/B})$.
This level peels away from the inner-mode continuum as $\abs{\gamma}$ increases further, eventually becoming a zero-energy state (independent of $\phi$) when $\abs{\gamma} = \pi/2$.

\begin{figure}[t]
  \begin{center}
    \includegraphics[width=0.35\textwidth]{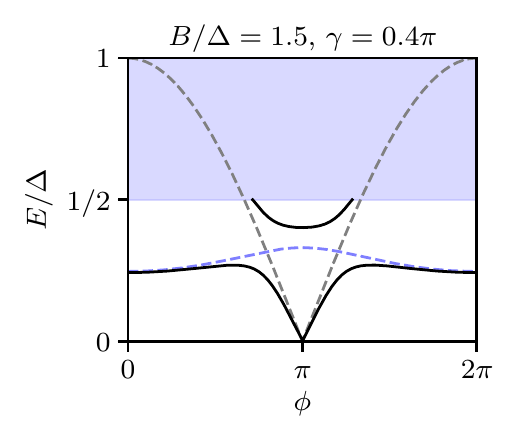}
    \caption{Single-particle excitation spectrum of the junction in the presence of a very large forward scattering phase ($\gamma = 0.4\pi$) in the topological regime, $B = 1.5\Delta$. In the limit $D=1$, since $\gamma > \arcsin(\sqrt{\Delta/B}) \approx 0.3\pi$, a bound state is present in the inner modes at all $\phi$ (dashed blue curve). At $D < 1$ this state hybridizes with the outer-mode bound state (dashed gray curve) resulting in one energy level that is separated from the continuum at all phase differences, and a second level that dips below the continuum only near $\phi = \pi$ (solid black curves, $D = 0.975$). The spectrum may be compared with that at $|\gamma|\ll 1$, cf.~Fig.~\ref{fig:andreev_spectrum_backscattering}(c), as well as with the spectrum at $\gamma \sim 1$ but $\gamma < \arcsin(\sqrt{\Delta/B})$, cf.~Fig.~\ref{fig:andreev_spectrum_fwd_scatt} (note, however, that those Figures are at different values of $B/\Delta$ and $D$ than this one).}
    \label{fig:andreev_spectrum_fwd_scatt2}
  \end{center}
\end{figure}

As discussed in Sec.~\ref{sec:scat} of the main text, weak backscattering, $1-D \ll 1$, may cause the inner-mode level(s) to hybridize with the outer-mode bound state.
Recall that, at $D=1$, the outer-mode bound state $E_0(\phi)$ reaches the edge of the inner-mode continuum at $\phi = \pi \pm \phi_0$, where $\phi_0$ is given by Eq.~\eqref{eq:phi0}, while the inner-mode shallow states exist for $\abs{\phi} < \phi_1$ and $\abs{\phi - 2\pi} < \phi_1$.
Hence if $\phi_0 + \phi_1 \ll \pi$ (i.e.~for sufficiently small $\abs{\gamma}$), the hybridization is ineffective, resulting in a spectrum similar to Fig.~\ref{fig:andreev_spectrum_backscattering}(c).
As $\abs{\gamma}$ increases past the point where $\phi_0 + \phi_1 \approx \pi$, the hybridization becomes effective, leading to a single energy level that is separated from the continuum at all $\phi$, as shown in Fig.~\ref{fig:andreev_spectrum_fwd_scatt}.
Using Eqs.~\eqref{eq:phi0} for $\phi_0$ and \eqref{eq:app_phi1} for $\phi_1$, the condition $\phi_0 + \phi_1 \gtrsim \pi$ is equivalent to
\begin{equation}
    \abs{\sin{\gamma}} \gtrsim \sqrt{1 - (B-\Delta)/\Delta} .
    \vspace{0em}
\end{equation}
Near the transition, $B - \Delta \ll \Delta$, this reduces to $\abs{\gamma \pm \pi/2} \lesssim \sqrt{(B-\Delta)/\Delta}$, i.e., the condition stated in Sec.~\ref{sec:bs_scatt_top} of the main text.

Finally, let us consider the effect of weak backscattering when $\abs{\gamma} > \arcsin(\sqrt{\Delta/B})$ is so large that the bound state in the inner modes is already separated from the continuum at all phases $\phi$.
Now hybridization with the outer-mode bound state may lead to the situation depicted in Fig.~\ref{fig:andreev_spectrum_fwd_scatt2}, in which a second energy level dips below the continuum at phases near $\phi = \pi$.\\

\subsection{Inner-mode continuum states with forward scattering}

At energies $E > \abs{\Delta - B}$, the determinant of the $S$-matrix of the inner modes is given by (see Appendix~\ref{app:imode_s_matrix})
\begin{equation}
\label{eq:det_Si_fwsc}
    \det S_\mathrm{i}(E,\phi,\gamma) = \frac{[\Lambda_\mathrm{i}(E,\phi,\gamma)]^\star}{\Lambda_\mathrm{i}(E,\phi,\gamma)},
\end{equation}
with $\Lambda_{\mathrm{i}}$, defined in Eq.~\eqref{eq:app_imode_Lambda}, analytically continued into the appropriate range of above-gap energies by taking $[(\Delta \pm B)^2 - E^2]^{1/2} \to -i [E^2 - (\Delta \pm B)^2]^{1/2}$.
The scattering-induced contribution to the continuum density of states is then given by Eq.~\eqref{eq:rho_S}: $\delta\rho = (2\pi i)^{-1} \partial_E \ln \det S_\mathrm{i}$ [since $\det S = (\det S_{\mathrm{i}}) \cdot (\det S_{\mathrm{o}})$ and $\det S_{\mathrm{o}} = 1$].
It vanishes for $E > \Delta + B$.
Explicit calculation for $|\Delta - B| < E < \Delta + B$ yields
\begin{widetext}
\begin{equation}
\label{eq:rho_i_fwsc}
    \delta \rho(E,\phi,\gamma) = \frac{\partial}{\partial E} \frac{1}{\pi} \arctan\!\left( \frac{(E^2 + \Delta^2 - B^2) F(\phi) [1+G(\phi,\gamma)] + (E^2 - \Delta^2 + B^2) G(\phi,\gamma)}{\sqrt{[E^2-(\Delta-B)^2][(\Delta+B)^2-E^2]}} \right) ,
\end{equation}
where $F(\phi)$ and $G(\phi,\gamma)$ are defined in Eq.~\eqref{eq:app_FG_def}.
Equations~\eqref{eq:det_Si_fwsc} and \eqref{eq:rho_i_fwsc} generalize Eqs.~\eqref{eq:det_Si} and \eqref{eq:rho_i} for $\det S_\mathrm{i}(E,\phi)$ and $\delta \rho(E,\phi)$ respectively, and reduce to the latter in the limit $\gamma \to 0$.

\clearpage

\section{Spectrum at nonzero backscattering, \texorpdfstring{$D < 1$}{D < 1}}
\label{app:full_spectrum}

We next determine the spectrum in the presence of backscattering, $D < 1$. 
To streamline notation, we write
\begin{equation}
    \Phi(x) = \begin{bmatrix}
    \Phi_\textrm{i}(x) \\
    \Phi_\textrm{o}(x)
    \end{bmatrix} ,
\end{equation}
and introduce new Pauli matrices $\chi_i$ acting in the space of inner ($\chi_z = + 1$) and outer ($\chi_z = - 1$) modes.
Performing the gauge transformation \eqref{eq:app_gauge_transf}, the linearized BdG equations \eqref{eq:BdG_equations} take the form
\begin{equation}
\label{eq:app_full_BdG}
    \big[ \! -i\alpha \pd_x \tau_z \sigma_z \chi_z - \tfrac{1}{2} B \sigma_x (1+\chi_z) + \Delta \tau_x \big] \Phi(x) 
    \equiv \mathcal{H}  \Phi(x) = E \Phi(x) ,
\end{equation}
where the first equality defines $\mathcal{H}$, the linearized BdG hamiltonian.
The boundary condition at $x=0$, Eq.~\eqref{eq:BdG_bc}, becomes
\begin{equation}
\label{eq:app_full_bc}
    \Phi(0^+)
    = e^{-i (\phi/2) \tau_z} \, \frac{e^{i\gamma \sigma_z \chi_z}}{\sqrt{D}} \left[ 1 + \sqrt{1-D} \, \chi_x \right] \Phi(0^-) 
    \equiv T(\phi,D,\gamma) \, \Phi(0^-) .
\end{equation}
Here the last equality defines the transfer matrix $T$ across the junction.
For convenience, we have set the reflection phase $\delta = 0$; as discussed in the main text, $\delta$ can be eliminated from the linearized BdG problem by a unitary transformation, so its actual value does not affect the spectrum of the junction in the limit we are considering.
The solutions of Eqs.~\eqref{eq:app_full_BdG}, \eqref{eq:app_full_bc} at energy $E$ are related to those at energy $-E$ by the anti-unitary ``particle-hole'' operator
\begin{equation}
\label{eq:app_full_ph}
    \mathcal{P} = \tau_y \sigma_y \mathcal{K} ,
\end{equation}
where $\mathcal{K}$ denotes the operator of complex conjugation. 
The operator $\mathcal{P}$ anticommutes with $\mathcal{H}$ and commutes with $T$, so 
the solutions at energies $\pm E$ are indeed mapped into one another by the action of $\mathcal{P}$.

From the analysis in Appendix~\ref{app:imode_spectrum}, it follows that the generalized plane-wave solutions of Eq.~\eqref{eq:app_full_BdG} at energy $E$ are $\Phi^{\mathrm{i}}_{\sigma,\nu}(E) e^{i \sigma k_{\nu}(E) x}$ and $\Phi^{\mathrm{o}}_{\sigma,\nu}(E) e^{-i \sigma k_0(E) x}$, where 
\begin{equation}
\label{eq:app_full_kE}
    k_{\nu}(E) = \frac{1}{\alpha} \, [E^2 - (\Delta + \nu B)^2]^{1/2} ,
\end{equation}
and where
\begin{subequations}
\label{eq:app_full_Phi}
\begin{align}
    \Phi^{\mathrm{i}}_{\sigma,+}(E) &= 
    \frac{1}{2\sqrt{i \sin{\eta_+(E)}}} \begin{bmatrix}
    e^{i\sigma \eta_+(E)/2} \\
    - e^{-i\sigma \eta_+(E)/2} \\
    e^{-i\sigma \eta_+(E)/2} \\
    - e^{i\sigma \eta_+(E)/2} \\
    \mathbf{0}_4
    \end{bmatrix} \! ,  \qquad
    \Phi^{\mathrm{i}}_{\sigma,-}(E) = 
    \frac{1}{2\sqrt{i \sin{\eta_-(E)}}} \begin{bmatrix}
    e^{i\sigma \eta_-(E)/2} \\
    \sgn(\Delta - B) \, e^{-i\sigma \eta_-(E)/2} \\
    \sgn(\Delta - B) \, e^{-i\sigma \eta_-(E)/2} \\
    e^{i\sigma \eta_-(E)/2} \\
    \mathbf{0}_4
    \end{bmatrix} \! , \\*
    \Phi^{\mathrm{o}}_{\sigma,+}(E) &= \,
    \frac{1}{2\sqrt{i \sin{\eta_0(E)}}} \begin{bmatrix}
    \mathbf{0}_4 \\
    e^{i\sigma \eta_0(E)/2} \\
    - e^{-i\sigma \eta_0(E)/2} \\
    e^{-i\sigma \eta_0(E)/2} \\
    - e^{i\sigma \eta_0(E)/2}
    \end{bmatrix} \! ,  \qquad \,\,
    \Phi^{\mathrm{o}}_{\sigma,-}(E) = 
    \frac{1}{2\sqrt{i \sin{\eta_0(E)}}} \begin{bmatrix}
    \mathbf{0}_4 \\
    e^{i\sigma \eta_0(E)/2} \\
    e^{-i\sigma \eta_0(E)/2} \\
    e^{-i\sigma \eta_0(E)/2} \\
    e^{i\sigma \eta_0(E)/2}
    \end{bmatrix} \! .
\end{align}
\end{subequations}
Here, $\mathbf{0}_4 = (0,0,0,0)^T$, and
\begin{equation}
\label{eq:app_full_etaE}
    e^{i\eta_{\nu}(E)} = \frac{E + \alpha k_{\nu}(E)}{\abs{\Delta + \nu B}} .
\end{equation}
As before, we choose branches so that $k_{\nu}(E)$ always lies in the upper-right quadrant of the complex $k$-plane ($\re k \geq 0$, $\im k \geq 0$), and so that $\eta_{\nu}(E)$ is a continuous function of $E$ that approaches $-i \infty$ as $E \to \infty$. Thus, $\eta_{\nu}(E)$ is purely imaginary for $E > \abs{\Delta + \nu B}$, and real for $\abs{E} < \abs{\Delta + \nu B}$.

Any function $\Phi(x;E)$ that satisfies Eq.~\eqref{eq:app_full_BdG} can be expressed as a linear combination of the generalized plane waves at that energy $E$:
\begin{equation}
\label{eq:app_full_exp}
    \Phi(x;E,\phi) = \begin{cases}
    a^{\mathrm{i}}_{-,\nu} \, \Phi^{\mathrm{i}}_{-,\nu} e^{-ik_{\nu} x} 
    + b^{\mathrm{i}}_{+,\nu} \, \Phi^{\mathrm{i}}_{+,\nu} e^{+ik_{\nu} x}
    + a^{\mathrm{o}}_{+,\nu} \, \Phi^{\mathrm{o}}_{+,\nu} e^{-ik_0 x} 
    + b^{\mathrm{o}}_{-,\nu} \, \Phi^{\mathrm{o}}_{-,\nu} e^{+ik_0 x} , &\quad x > 0 , \\[0.1em]
    a^{\mathrm{i}}_{+,\nu} \, \Phi^{\mathrm{i}}_{+,\nu} e^{+ik_{\nu} x} 
    + b^{\mathrm{i}}_{-,\nu} \, \Phi^{\mathrm{i}}_{-,\nu} e^{-ik_{\nu} x}
    + a^{\mathrm{o}}_{-,\nu} \, \Phi^{\mathrm{o}}_{-,\nu} e^{+ik_0 x} 
    + b^{\mathrm{o}}_{+,\nu} \, \Phi^{\mathrm{o}}_{+,\nu} e^{-ik_0 x} , &\quad x < 0 ,
    \end{cases}
\end{equation}
where a summation over $\nu = \pm$ is implicit, $k_{\nu}(E)$, $\Phi^{\chi}_{\sigma,\nu}(E)$ ($\chi = \mathrm{i},\mathrm{o}$) are given in Eqs.~\eqref{eq:app_full_kE}, \eqref{eq:app_full_Phi} respectively, and we have suppressed all arguments for brevity.
When the wavevectors are real, the coefficients $a^{\chi}_{\sigma,\nu}$ ($b^{\chi}_{\sigma,\nu}$) multiply incoming (outgoing) plane waves.
When the wavevectors are imaginary, $a^{\chi}_{\sigma,\nu}$ ($b^{\chi}_{\sigma,\nu}$) multiply the generalized plane waves that grow (decay) exponentially at infinity.

The boundary condition at $x=0$, Eq.~\eqref{eq:app_full_bc}, then yields a linear system of equations to be satisfied by the coefficients:
\begin{equation}
\label{eq:app_full_bc_ab}
    \Omega_a(E,\phi,D,\gamma) \, \vec{a} = \Omega_b(E,\phi,D,\gamma) \, \vec{b} ,
\end{equation}
where
\begin{subequations}
\begin{align}
    \vec{a} &= \big( a^{\mathrm{i}}_{++}, \, a^{\mathrm{i}}_{+-}, \,
    a^{\mathrm{o}}_{-+}, \, a^{\mathrm{o}}_{--}, \,
    a^{\mathrm{i}}_{-+}, \, a^{\mathrm{i}}_{--}, \,
    a^{\mathrm{o}}_{++}, \, a^{\mathrm{o}}_{+-} \big)^T , \\
    \vec{b} &= \big( b^{\mathrm{i}}_{-+}, \ b^{\mathrm{i}}_{--}, \
    b^{\mathrm{o}}_{++}, \ b^{\mathrm{o}}_{+-}, \,
    b^{\mathrm{i}}_{++}, \ b^{\mathrm{i}}_{+-}, \ 
    b^{\mathrm{o}}_{-+}, \ b^{\mathrm{o}}_{--} \big)^T ,
\end{align}
\end{subequations}
and $\Omega_a, \Omega_b$ are $8 \times 8$ matrices, given by
\begin{subequations}
\begin{align}
\label{eq:app_full_OmegaA}
    \Omega_a(E,\phi,D,\gamma) &= \Big[ 
    T \Phi^{\mathrm{i}}_{+,+}, \ 
    T \Phi^{\mathrm{i}}_{+,-}, \ 
    T \Phi^{\mathrm{o}}_{-,+}, \ 
    T \Phi^{\mathrm{o}}_{-,-}, \
    -\Phi^{\mathrm{i}}_{-,+}, \ 
    -\Phi^{\mathrm{i}}_{-,-}, \
    -\Phi^{\mathrm{o}}_{+,+}, \ 
    -\Phi^{\mathrm{o}}_{+,-} \Big] , \\*[0.2em]
    \label{eq:app_full_OmegaB}
    \Omega_b(E,\phi,D,\gamma) &= \Big[ 
    \! - \! T \Phi^{\mathrm{i}}_{-,+}, \ 
    -T \Phi^{\mathrm{i}}_{-,-}, \
    -T \Phi^{\mathrm{o}}_{+,+}, \ 
    -T \Phi^{\mathrm{o}}_{+,-}, \
    \Phi^{\mathrm{i}}_{+,+}, \ 
    \Phi^{\mathrm{i}}_{+,-}, \ 
    \Phi^{\mathrm{o}}_{-,+}, \ 
    \Phi^{\mathrm{o}}_{-,-} \Big] . 
\end{align}
\end{subequations}
As before, $[\Phi, \Psi, \cdots]$ denotes the matrix whose first column is $\Phi$, second column $\Psi$, and so on. $T =  T(\phi,D,\gamma)$ is the transfer matrix across the junction, defined in Eq.~\eqref{eq:app_full_bc} above.
The wavefunction $\Phi(x;E,\phi)$ in Eq.~\eqref{eq:app_full_exp} represents a valid solution of the linearized BdG problem if and only if (i) the coefficients $a^{\chi}_{\sigma,\nu}, b^{\chi}_{\sigma,\nu}$ satisfy Eq.~\eqref{eq:app_full_bc_ab}, and (ii) $\Phi(x;E,\phi)$ satisfies appropriate conditions at $x = \pm \infty$.

\subsection{Bound states}
\label{app:full_bound_state}

We first consider states belonging to the discrete spectrum (bound states), which correspond to normalizable wavefunctions $\Phi(x;E,\phi)$. Normalizability requires that the wavevectors $\sigma k_{\nu}(E)$ appearing in the expansion \eqref{eq:app_full_exp} all have positive (negative) imaginary part for $x > 0$ ($x < 0$).
A similar analysis to that in Appendix~\ref{app:imode_bound_state} shows that bound states can only exist at energies $\abs{E} < \min(\abs{\Delta - B},\Delta)$, where all the $k$'s are imaginary.
The bound state wavefunction then has the form \eqref{eq:app_full_exp} with $a^{\chi}_{\sigma,\nu} \equiv 0$. The $b^{\chi}_{\sigma,\nu}$ coefficients must still satisfy Eq.~\eqref{eq:app_full_bc_ab}, which reduces to:
\begin{equation}
    0 = \Omega_b(E,\phi,D,\gamma) \, \vec{b} .
\end{equation}
This equation has a nontrivial solution for $\vec{b}$ (and hence a bound state exists at the given $E$ and $\phi$) if and only if $\det{\Omega_b} = 0$.
Explicit calculation using Eqs.~\eqref{eq:app_full_Phi} and \eqref{eq:app_full_OmegaB} yields
\begin{equation}
\label{eq:app_full_det_OmegaB}
    \det{\Omega_b} = \frac{1}{2\Delta^2(\Delta^2-B^2) \, \sin^2\!\eta_0 \, \sin{\eta_-} \sin{\eta_+}} \, \frac{\Lambda(E,\phi;D,\gamma)}{D^2} ,
\end{equation}
where
\begin{equation}
\label{eq:app_full_Lambda}
    \Lambda(E,\phi;D,\gamma) 
    = \Lambda_0(E,\phi) + (1-D) \Lambda_1(E,\phi) + (1-D)^2 \Lambda_2(E,\phi) + (D \sin^2\!\gamma) \, \Lambda_{\gamma}(E,\phi) ,
\end{equation}
with
\begin{subequations}
\label{eq:app_full_Lambda_parts}
\begin{align}
    \Lambda_0(E,\phi) 
    = \, &\big[ \Delta^2 \cos^2(\phi/2) - E^2 \big] 
    \big[\! -\alpha k_- \alpha k_+ (1+\cos^2(\phi/2)) - (E^2 + \Delta^2 - B^2) \sin^2(\phi/2) \big] , \\*[0.5em]
    \Lambda_1(E,\phi) 
    = \, &\big[ (E^2 + \Delta^2 - B^2) (2\Delta^2\cos^2(\phi/2) - E^2) - 2\Delta^2E^2 \big] \sin^2(\phi/2) \nonumber \\*
    &- \alpha k_0 \alpha k_- \big[ \Delta (\Delta + B) \cos^2(\phi/2) - E^2 \big] - \alpha k_0 \alpha k_+ \big[ \Delta (\Delta - B) \cos^2(\phi/2) - E^2 \big] \nonumber \\*
    &+ \alpha k_- \alpha k_+ \big[ 2\Delta^2\cos^4(\phi/2) - E^2 (1+\cos^2(\phi/2)) \big] , \\*[0.5em]
    \Lambda_2(E,\phi) 
    = \, &\Delta^2 (\Delta^2 - B^2 + E^2) \sin^4(\phi/2) 
    + \tfrac{1}{2} (\Delta^2 - B^2 - E^2) \big[ \Delta^2 \cos{\phi} - E^2 \big] - \tfrac{1}{2} B^2 E^2 \nonumber \\*
    &+ \tfrac{1}{2} \alpha k_0 \alpha k_- \big[ \Delta (\Delta + B) \cos{\phi} - E^2 \big] 
    + \tfrac{1}{2} \alpha k_0 \alpha k_+ \big[ \Delta (\Delta - B) \cos{\phi} - E^2 \big] \nonumber \\*
    &- \tfrac{1}{2} \alpha k_- \alpha k_+ \big[ \Delta^2(1 - \tfrac{1}{2} \sin^2\!\phi) - E^2 \big] , \\*[0.5em]
    \Lambda_\gamma(E,\phi) 
    = \, &\big[ \Delta^2 \cos^2(\phi/2) - E^2 \big] \big[ \alpha k_- \alpha k_+ + \Delta^2 - B^2 - E^2 \big] ,
\end{align}
\end{subequations}
and
\begin{equation}
\label{eq:app_k_nuE}
    \alpha k_{\nu}(E) = i \, [(\Delta + \nu B)^2 - E^2]^{1/2} .
\end{equation}
Note that Eq.~\eqref{eq:app_full_Lambda} above reduces to $\Lambda_\mathrm{i}(E,\phi) \cdot \Lambda_{\mathrm{o}, +}(E,\phi) \cdot \Lambda_{\mathrm{o}, -}(E,\phi)$ [see Eqs.~\eqref{eq:Lambda_o}, \eqref{eq:Lambda_i}] in the limit $D = 1$, $\gamma = 0$ (up to an overall multiplicative real, $E$-independent factor; the latter does not influence the spectrum of the junction).

The bound state energies $E^\mathrm{sc}_i(\phi)$ (where the index $i$ ennumerates the bound states) are thus solutions of the equation $\Lambda(E,\phi;D,\gamma) = 0$ in the energy interval $\abs{E} < \min(\abs{\Delta - B},\Delta)$.
They can be obtained analytically only in particular limits, analyzed in detail in Appendix~\ref{app:approx_spectrum}.
In general, it is possible to determine the bound state energies $E^\mathrm{sc}_i(\phi)$ numerically.
To do so, we evaluate $\Lambda(E,\phi;D,\gamma)$ on a fine grid of $E$ values spanning the range $[0, \, \min(\abs{\Delta - B},\Delta)]$, identify all intervals in which the function changes sign, and then apply a stable root-finding algorithm (Brent's method) within each sign-changing interval.

\subsection{Scattering matrix}
\label{app:full_s_matrix}

We next consider states belonging to the continuous spectrum (scattering states). 
For any $E$ and $\phi$ such that $\det \Omega_b(E,\phi,D,\gamma) \neq 0$, we may solve Eq.~\eqref{eq:app_full_bc_ab} to express $\vec{b}$ as a linear function of $\vec{a}$:
\begin{equation}
    \vec{b} = [\Omega_b(E,\phi,D,\gamma)]^{-1} \, \Omega_a(E,\phi,D,\gamma) \, \vec{a} .
\end{equation}
The scattering states are then given by Eq.~\eqref{eq:app_full_exp}. In order for the wavefunctions to remain finite as $x \to \pm \infty$, the coefficient $a^{\chi}_{\sigma,\nu}(E)$ must vanish unless the corresponding wavevector $k_{\nu}(E)$ is real. This condition fixes the number of linearly independent scattering states in the continuum at energy $E$.

The scattering matrix $S$ at energy $E$ is given by the block of $[\Omega_b(E,\phi,D,\gamma)]^{-1} \, \Omega_a(E,\phi,D,\gamma)$ corresponding to the modes that are propagating at that energy:
\begin{equation}
\label{eq:app_full_Smat}
    S(E,\phi,D,\gamma) = \big( [\Omega_b(E,\phi,D,\gamma)]^{-1} \, \Omega_a(E,\phi,D,\gamma) \big)_{\text{prop}} .
\end{equation}
Performing the matrix inversion and multiplication, we obtain an exact analytical expression for $S(E,\phi,D,\gamma)$.
This expression is quite unwieldy at $D < 1$, so we omit it here.
The determinant of the scattering matrix equals
\begin{equation}
\label{eq:app_full_detS}
    \det S(E,\phi,D,\gamma) = \frac{[\Lambda(E,\phi;D,\gamma)]^\star}{\Lambda(E,\phi;D,\gamma)} ,
\end{equation}
where ${}^\star$ denotes complex conjugation, and the function $\Lambda(E,\phi;D,\gamma)$ was defined in Eqs.~\eqref{eq:app_full_Lambda} and \eqref{eq:app_full_Lambda_parts}; it is analytically continued past energy thresholds by using the appropriate complex values for the wavevectors:
\begin{equation}
\label{eq:app_analytic_c}
    \alpha k_{\nu}(E) = \begin{cases}
    [E^2 - (\Delta + \nu B)^2]^{1/2} &\text{if} \quad E > \abs{\Delta + \nu B} , \\*
    i \, [(\Delta + \nu B)^2 - E^2]^{1/2} &\text{if} \quad E < \abs{\Delta + \nu B} .
    \end{cases}
\end{equation}

\clearpage

\section{Approximate results on spectrum of the junction in the presence of scattering}
\label{app:approx_spectrum}

In this Appendix we describe in detail how the approximate results on the spectrum of the system in the presence of scattering, which are presented in Sec.~\ref{sec:states_scat}, can be obtained.

\subsection{Hybridization of the bound states at \texorpdfstring{$B < \Delta$}{B < Δ}}

First, we discuss how the hybridization of the bound states, which happens near $\phi = \pi$ in the trivial phase, $B < \Delta$, can be described quantitatively by approximately solving the equation for the bound state energies, $\Lambda(E,\phi) = 0$. To begin with, we simplify the general expression for $\Lambda(E,\phi)$, Eq.~\eqref{eq:app_full_Lambda}, under the assumptions
\begin{equation}
\label{eq:app_condition}
    |E| \ll \Delta - B \ll \Delta, \qquad 
    |\phi - \pi| \ll 1, \qquad 
    1-D \ll \frac{\Delta - B}{\Delta}, \qquad 
    |\gamma|\ll 1.
\end{equation}
To leading order in the small parameters we find
\begin{equation}
\label{eq:app_lowest}
    \Lambda(E,\phi)\approx \frac{\Delta}{\Delta-B}
    \Big[ (E_1^2 - E^2)\left(E_0^2 - E^2\right) -4 (1-D)(\Delta - B) \Delta\left(E^2 - E_0 E_1\right)+4(1-D)^2\Delta^2 (\Delta - B)^2 \Big].
\end{equation}
Here $E_0(\phi) \approx -\Delta(\phi-\pi)/2$, $E_1(\phi)\approx -(\Delta - B)(\phi - \pi)$ are the expressions for the energies of outer and inner-mode bound states [Eqs.~\eqref{eq:E0} and \eqref{eq:bs_energy}], evaluated at
$|\phi - \pi| \ll 1$.
Solving the equation $\Lambda(E,\phi) = 0$, we obtain the bound state energies $E^\mathrm{sc}_{0,1}(\phi)$ that are given by Eq.~\eqref{eq:bs_hybr}.

Eq.~\eqref{eq:bs_hybr} indicates that the two energy levels, $E^\mathrm{sc}_0(\phi)$ and $E^\mathrm{sc}_1(\phi)$, cross at $\phi = \pi$. This peculiarity of the lowest-order calculation is disrupted by a small forward scattering phase $|\gamma|\ll 1$. The latter produces a subleading correction $\delta\Lambda = - \Delta \gamma^2 E^4/(\Delta - B)$ to Eq.~\eqref{eq:app_lowest}, which results in an anticrossing between the Andreev levels at $\phi = \pi$. By taking this correction into account, from $\Lambda = 0$ we find the bound state energies $E^\mathrm{sc}_{\pm}(\phi)$ presented in Eq.~\eqref{eq:anticross}. Equation~\eqref{eq:anticross} indicates that the anticrossing is manifested in a narrow vicinity $\sim |\gamma|\,\delta\varepsilon /\Delta \ll 1$ of $\phi = \pi$ only [recall, that $\delta\varepsilon$ describes the separation of the bound states' energies from zero at $\phi = \pi$ and is given by Eq.~\eqref{eq:deltaep}]. It is characterized by an energy splitting $E_+^\mathrm{sc}(\pi) - E_-^\mathrm{sc}(\pi) = |\gamma| \delta \varepsilon \ll \delta \varepsilon$.

The anticrossing relies on the presence of a forward scattering phase. In the case $\gamma = 0$, the Andreev levels cross at $\phi = \pi$ even beyond the accuracy of Eq.~\eqref{eq:app_lowest}. To see this, notice that the following relation between $\Lambda_0$, $\Lambda_1$, and $\Lambda_2$ holds at $\phi = \pi$ [see Eq.~\eqref{eq:app_full_Lambda_parts}]:
\begin{equation}
    4\Lambda_0(E, \pi)\Lambda_2(E, \pi) = \Lambda_1^2(E, \pi).
\end{equation}
Thus, at $\gamma = 0$, the expression \eqref{eq:app_full_Lambda} for $\Lambda(E,\pi)$ is a complete square; consequently, each energy level at $\phi = \pi$ is two-fold degenerate.

\subsection{Hybridization between the outer-mode bound state and the states of the spectral continuum in the topological phase} 
\label{sec:app_top_bs}

Next, we employ the general expression for $\Lambda(E,\phi)$, Eq.~\eqref{eq:app_full_Lambda}, to describe how the energy of the outer-mode bound state, $E_0^\mathrm{sc}(\phi)$, is affected by scattering in the topological phase, $B > \Delta$, near the continuum edge. 
In that, we focus on the vicinity of the topological transition and on phases $\phi$ close to $\pi$,
\begin{equation}
    0 < B-\Delta \ll \Delta, \qquad 
    |\phi-\pi| \ll 1.
\end{equation}
Furthermore, we assume that the scattering is weak,
\begin{equation}
    1-D\ll 1,\qquad 
    |\gamma|\ll 1.
\end{equation}
Given these approximations, Eq.~\eqref{eq:app_full_Lambda} can be simplified considerably at $|E| \ll \Delta$. Performing a lowest-order expansion in small parameters, we find
\begin{equation}
\label{eq:app_lambda_top_simpl}
    \Lambda (E,\phi) \approx 2 \Delta (B - \Delta)
    \Bigg[ E_{0}^{2}-E^{2}-2\Delta(1-D)\frac{E^2}{B-\Delta}-\Delta^{2}(1-D)^{2} + \sqrt{1-\frac{E^{2}}{(B-\Delta)^{2}}}\left(E_{0}^{2}-E^{2}+\Delta^{2}(1-D)^{2}\right) \Bigg].
\end{equation}
Here $E_0(\phi) \approx -\Delta (\phi - \pi) / 2$ is the energy of the outer-mode bound state in a transparent junction. 

By solving $\Lambda(E,\phi) = 0$ for the phase $\phi$ at $|E| = B - \Delta$, we establish that, in the presence of scattering, the energy $E_0^\mathrm{sc}$ of the outer-mode bound state reaches the continuum edge at points $\phi = \pi \pm \phi_0^\mathrm{sc}$, where
\begin{equation}
\label{eq:app_phi0sc}
    \phi_{0}^{\mathrm{sc}}	\approx 2\frac{B-\Delta}{\Delta}+2(1-D).
\end{equation}
The behavior of $E_0^\mathrm{sc}(\phi)$ near these points can be addressed concisely in the limit of perturbatively weak backscattering, $1 - D \ll (B - \Delta)/\Delta$. By taking
\begin{equation}
    \delta\phi = \pi + \phi_0^\mathrm{sc} - \phi \ll \phi_0^\mathrm{sc}, \qquad
    \delta E = B - \Delta + E \ll B - \Delta
\end{equation}
[notice that we assume $-(B-\Delta) < E < 0$ here], we obtain the following approximate expression:
\begin{equation}
    \Lambda(E, \phi) \approx 4\Delta (B-\Delta)^2  \left[ \delta E+\Delta(1 - D) \sqrt{2\,\delta E/(B-\Delta)}-\Delta \delta\phi/2 \right].
\end{equation}
Then, by solving the equation for the bound state energies, $\Lambda = 0$, we arrive at expression \eqref{eq:boundsc} for $E_0^\mathrm{sc}(\phi)$.

A small forward scattering phase, $|\gamma|\ll 1$, gives only a subleading correction to Eq.~\eqref{eq:app_lambda_top_simpl}. This correction has a form
\begin{equation}
    \delta \Lambda (E,\phi) \approx
    -2\Delta(B-\Delta)\gamma^{2}\big(E_{0}^{2}-E^{2}\big)\Bigl(\sqrt{1-E^{2}/(B-\Delta)^{2}}+1\Bigr).
\end{equation}
It results in a shift of the points where the bound state reaches the continuum edge from $\pi \pm \phi_0^\mathrm{sc}$ to $\pi + \phi_0^\mathrm{sc,\gamma}$, where [for $1 - D \ll (B - \Delta)/\Delta$]
\begin{equation}
    \phi_0^\mathrm{sc,\gamma} \approx 2\frac{B-\Delta}{\Delta} + 2(1-D)(1+\gamma^2/2).
\end{equation}
This differs from $\phi^\mathrm{sc}_0$ [see Eq.~\eqref{eq:app_phi0sc}] by a slight modification of the numeric prefactor in front of $1 - D$ only. Such prefactors experience the same modification in the expression for the bound state energy.
For $\delta \phi = \pi + \phi_0^\mathrm{sc,\gamma} - \phi \ll \phi_0^\mathrm{sc,\gamma}$ we find [cf.~Eq.~\eqref{eq:boundsc}]
\begin{equation}
    E_0^\mathrm{sc,\gamma}(\phi) \approx - (B - \Delta) 
    \Bigg\{ 1 - \frac{\Delta^2}{2(B-\Delta)^2} \bigg[\sqrt{\bigl[(1-D)(1+\gamma^2/2)\bigr]^2+\frac{B-\Delta}{\Delta}\delta\phi} - (1-D)(1+\gamma^2/2)\bigg]^2 \Bigg\} .
\end{equation}

\subsection{Shallow bound states}
\label{app:sbs}

As a next step, we use Eq.~\eqref{eq:app_full_Lambda} to approximately find the energies of the shallow bound states, $E_\mathrm{s}^\mathrm{sc}(\phi)$. As stated in Sec.~\ref{sec:scat}, such states appear near $\phi = 0,\:2\pi$ in the topological phase in the presence of weak scattering, $1-D,|\gamma|\ll 1$. 
First, we compute the function $\Lambda(E,\phi)$ at energies $B - \Delta - E \ll B - \Delta$ in the vicinity of the topological transition, $B - \Delta \ll \Delta$. We focus on phases $0\leq \phi \ll 1$ [the energy of the bound state near $\phi = 2\pi$ can be obtained through $E_\mathrm{s}^\mathrm{sc}(2\pi - \phi) = E_\mathrm{s}^\mathrm{sc}(\phi)$]. Performing an expansion of $\Lambda$ to leading order in $(B-\Delta)/\Delta$, $\phi$, $1-D$, $\gamma$, and $(B - \Delta - E)/(B-\Delta)$, we get
\begin{equation}
\label{eq:Lambda_shallow}
    \Lambda(E,\phi)\approx 2\Delta^3(B-\Delta) \left\{2\sqrt{2\Bigl(1-\frac{E}{B-\Delta}\Bigr)} - \Bigl[ 1-D + \gamma^2 - \frac{\phi^2}{4} \Bigr]\right\}.
\end{equation}
This expression demonstrates that the equation for the bound state energies, $\Lambda(E,\phi) = 0$, admits a solution $E_\mathrm{s}^\mathrm{sc}(\phi)$ in the interval $\phi \in \left[0, \phi_\mathrm{s}\right)$ only (for $0\leq \phi \ll 1$), where
$\phi_\mathrm{s} = 2\sqrt{1 - D +\gamma^2}$. The resulting $E_\mathrm{s}^\mathrm{sc}(\phi)$ is given by Eq.~\eqref{eq:shallow}. Notice that the weakness of the scattering implies $B - \Delta - E_\mathrm{s}^\mathrm{sc}(0)\sim (B-\Delta) (1-D + \gamma^2)^2 \ll (B-\Delta)$ and $\phi_\mathrm{s}\ll 1$. These inequalities validate the applicability of the lowest-order expansion which was used to derive Eq.~\eqref{eq:Lambda_shallow}.

\clearpage

\end{widetext}

\section{Josephson current in the presence of scattering}
\label{app:approx_scat_jc}

In this Appendix we describe a convenient approach to the calculation of the Josephson current (Sec.~\ref{app:jc}), which was used to produce Figs.~\ref{fig:current_bs} and \ref{fig:strong_bs}.  
Additionally, we discuss how the approximate expressions for the Josephson current, Eqs.~\eqref{eq:trivialIsm} and \eqref{eq:I_top}, were obtained (Sec.~\ref{app:jc_appr}).

\subsection{General expressions for \texorpdfstring{$I(\phi)$}{I(ϕ)}} 
\label{app:jc}

\subsubsection{Trivial phase, \texorpdfstring{$B < \Delta$}{B < Δ}}
\label{app:jc_triv}

We first consider the junction in the trivial phase, $B < \Delta$, and discuss how the Josephson current $I(\phi)$ can be expressed directly in terms of $\Lambda(E,\phi)$. As a first step, we divide $I(\phi)$ [which is related to the ground state energy of the junction through Eq.~\eqref{eq:JC}] into two contributions,
\begin{equation}
    I(\phi) = I^{(1)}(\phi) + I^{(2)}(\phi).
\end{equation}
The first contribution, $I^{(1)}(\phi)$, originates from the Andreev bound states. The second contribution, $I^{(2)}(\phi)$, comes from the states above the continuum edge, $E > \Delta - B$. 

\begin{figure}[t]
  \begin{center}
    \includegraphics[width=0.4\textwidth]{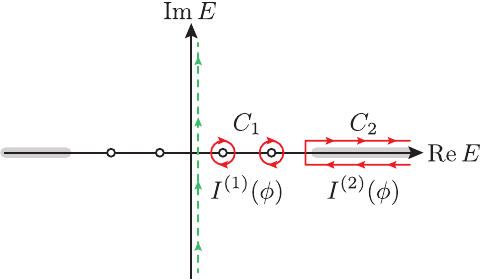}
    \caption{Integration contours $C_1$ and $C_2$ in the complex plane of energy that are featured in the expressions for $I^{(1)}(\phi)$ [Eq.~\eqref{eq:app_I1}] and $I^{(2)}(\phi)$ [Eq.~\eqref{eq:app_I2}], respectively. Empty circles on the real axis correspond to the bound state energies. Grey thick lines on the real axis depict branch cuts of $\Lambda(E,\phi)$ at $|E| > |\Delta - B|$. The combined contour $C_1 \cup C_2$ can be deformed into a contour running along the imaginary axis, depicted with a dashed line.}
    \label{fig:contours}
  \end{center}
\end{figure}

In the trivial phase (and in the presence of scattering) all bound states have positive energies, $E^\mathrm{sc}_\mathrm{b}(\phi) > 0$ [index ``$\mathrm{b}$'' enumerates the bound states]. Therefore, these levels are not occupied in the ground state and the contribution $I^{(1)}(\phi)$ is given by
\begin{equation}
\label{eq:app_I1_exp}
    I^{(1)}(\phi) = -e\sum_\mathrm{b} \partial_\phi E^\mathrm{sc}_\mathrm{b}(\phi).
\end{equation}
The sum over the bound states in Eq.~\eqref{eq:app_I1_exp} can be expressed explicitly in terms of the function  $\Lambda(E,\phi)$. Indeed, the energies of the bound states correspond to zeros of $\Lambda(E,\phi)$ in the complex plane of $E$ (see discussion in Sec.~\ref{sec:scat}). Thus, $I^{(1)}(\phi)$ can be represented as the following contour integral:
\begin{equation}
\label{eq:app_I1}
    I^{(1)}(\phi) = e\int_{C_1} \frac{EdE}{2\pi i} \frac{\partial^2}{\partial E \partial \phi} \ln \Lambda (E,\phi).
\end{equation}
Here $C_1$ is a combination of contours that encircle all of $E^\mathrm{sc}_\mathrm{b}$ (see Fig.~\ref{fig:contours}). Next, we consider the contribution to the Josephson current due to the states with $E > \Delta - B$.  By using Eqs.~\eqref{eq:rho_S} and \eqref{eq:detS_lambda} we find
\begin{equation}
\label{eq:app_cont}
    I^{(2)}(\phi) = -e\int_{\Delta - B}^{+\infty}\frac{EdE}{2\pi i} \frac{\partial^2}{\partial E \partial \phi} \ln \frac{\Lambda^\star(E,\phi)}{\Lambda(E,\phi)}.
\end{equation}
The function $\Lambda(E, \phi)$ has a branch cut at $E > \Delta - B$. Expressions for $\Lambda(E,\phi)$ on two sides of the cut are related by complex conjugation, $\Lambda(E+i0,\phi) = \Lambda^\star (E-i0,\phi)$ [see Eq.~\eqref{eq:app_full_Lambda}]. This allows us to represent Eq.~\eqref{eq:app_cont} as a contour integral,
\begin{equation}
\label{eq:app_I2}
    I^{(2)}(\phi) = e \int_{C_2} \frac{EdE}{2\pi i} \frac{\partial^2}{\partial E \partial \phi} \ln \Lambda(E,\phi),
\end{equation}
where the contour $C_2$ is depicted in Fig.~\ref{fig:contours}. Here we assume that  relations \eqref{eq:app_analytic_c} are valid at the upper side of the cut.

The two contributions to the Josephson current, Eqs.~\eqref{eq:app_I1} and \eqref{eq:app_I2}, can be combined into a single expression by unfolding the unified integration contour $C_1 \cup C_2$ to the imaginary axis (see dashed contour in Fig.~\ref{fig:contours}). By doing this and then integrating by parts we obtain
\begin{equation}
\label{eq:app_J_triv}
    I(\phi) = -\frac{e}{2\pi} \int_{-\infty}^{+\infty} d\mathcal{E}\, \partial_\phi \ln \Lambda(i\mathcal{E},\phi).
\end{equation}
This expression is well-suited for a numeric calculation of $I(\phi)$ at any $B \leq \Delta$, $D$, and $\gamma$. We used it to produce Figs.~\ref{fig:current_bs}(a), (b).

We note that unfolding of the integration contour is well-justified mathematically. First, $\Lambda(E,\phi)$ does not have any zeros away from the real axis in the complex plane of $E$ for the branch choice that we use [see relations \eqref{eq:app_analytic_c} that are valid at the upper side of the positive-energy cut]. The latter corresponds to the ``physical sheet'' of the variable $E$ \cite{landaulifshitz3}. Additionally, the integrand $\partial_\phi \ln \Lambda$ falls off like $E^{-2}$ at $\abs{E} \to \infty$. Therefore, the semicircular contour at infinity does not contribute to $I(\phi)$.

\subsubsection{Topological phase, \texorpdfstring{$B > \Delta$}{B > Δ}}

Next, we obtain an expression for the Josephson current in the topological phase, assuming that the fermion parity of the junction is even.

To begin with, we note that at $\phi \in [-\pi,\pi]$ the Andreev bound state at the junction has positive energy $E^\mathrm{sc}_0(\phi) > 0$. Consequently, it is not occupied in the ground state. Then, following reasoning similar to that in Sec.~\ref{app:jc_triv}, we conclude that at $|\phi|\leq \pi$ the Josephson current can be computed through Eq.~\eqref{eq:app_J_triv}.

On the other hand, Eq.~\eqref{eq:app_J_triv} is not directly applicable at $|\phi| \in  (\pi, 2\pi]$. Across $\phi = \pi$ the energy of the Andreev level $E^\mathrm{sc}_0(\phi)$ becomes negative and the parity of the \textit{global} ground state switches from even to odd [see Fig.~\ref{fig:many-body_spectrum_scattering}]. In this case, Eq.~\eqref{eq:app_J_triv} gives the Josephson current in the odd fermion parity sector. To get  $I(\phi)$ in the even parity sector, Eq.~\eqref{eq:app_J_triv} has to be modified to account for the difference in the bound state occupation between even and odd states. Then, for $|\phi|\in (\pi, \pi + \phi_0^\mathrm{sc})$ (i.e., in the interval where the bound state is below the continuum edge) we find
\begin{equation}
\label{eq:app_current_top}
    I(\phi) = -\frac{e}{2\pi} \int_{-\infty}^{+\infty} d\mathcal{E}\,\partial_\phi \ln \Lambda(i\mathcal{E},\phi) - 2e \partial_\phi E^\mathrm{sc}_0(\phi).
\end{equation}
At $|\phi|\in [\pi + \phi_0^\mathrm{sc}, 2\pi - \phi_\mathrm{s}]$ the energy of the bound state $E^\mathrm{sc}_0(\phi)$ should be substituted in Eq.~\eqref{eq:app_current_top} by $-(B-\Delta)$. At $\phi \in (2\pi - \phi_\mathrm{s},2\pi]$ it should be replaced by the energy of the shallow bound state, $-E^\mathrm{sc}_\mathrm{s}(\phi)$. The resulting expression was used to produce Figs.~\ref{fig:current_bs}(c) and \ref{fig:strong_bs}.

\subsection{Approximate expression for the Josephson current}
\label{app:jc_appr}

\subsubsection{Estimate for the continuum contribution \texorpdfstring{$I^{(2)}(\phi)$}{I2(ϕ)} near \texorpdfstring{$\phi = \pi$}{ϕ = π} in the trivial phase, \texorpdfstring{$B < \Delta$}{B < Δ}}
\label{app:I2est}

Obtaining an approximate expression for $I(\phi)$ near $\phi = \pi$, Eq.~\eqref{eq:trivialIsm}, we calculated the contribution to the current due to the bound states $I^{(1)}(\phi)$ (with the help of Eq.~\eqref{eq:anticross}) and disregarded the continuum contribution $I^{(2)}(\phi)$. Here we confirm the validity of this approach for $|\phi - \pi|\ll 1$, $\Delta (1-D) \ll \Delta - B \ll \Delta$, $|\gamma|\ll 1$.

To begin with, we note that both $I^{(1)}(\phi)$ and $I^{(2)}(\phi)$ vanish at $\phi = \pi$ and deviate from zero linearly away from this point. For $|\phi - \pi| \lesssim [(1 - D)(\Delta - B)/\Delta]^{1/2}$ the contribution due to the bound states can be estimated as [see Eq.~\eqref{eq:trivialIsm}]
\begin{equation}
\label{eq:app_est_I1}
    I^{(1)}(\phi) \sim e\Delta (\pi - \phi) \left(\frac{\Delta}{\Delta - B}\right)^{1/2}(1-D)^{-1/2}.
\end{equation}
To calculate the continuum contribution $I^{(2)}(\phi)$ we expand $\Lambda (E,\phi)$ to second order in $(\phi - \pi)$ and then estimate the integral in Eq.~\eqref{eq:app_cont}. As a result, we find
\begin{equation}
\label{eq:app_est_I2}
    I^{(2)}(\phi) \sim e\Delta (\pi - \phi) \max \Bigl\{1, \frac{\Delta^2}{(\Delta - B)^2}(1-D)\Bigr\}.
\end{equation}
This estimate is valid as long as $|\phi - \pi| \ll (\Delta - B) / \Delta$.
Eqs.~\eqref{eq:app_est_I1} and \eqref{eq:app_est_I2} indicate that
\begin{equation}
\label{eq:app_ratio}
    I^{(2)}(\phi)/I^{(1)}(\phi) \ll 1
\end{equation}
within the whole phase interval $\delta\phi_{B<\Delta} \sim [(1 - D)(\Delta - B)/\Delta]^{1/2}$ around $\phi = \pi$, in which the smearing of the disconitnuity happens. This is a result of the condition $1-D \ll (\Delta - B)/\Delta$, which physically means that the energies of the bound states $\sim \delta \varepsilon = \sqrt{2\Delta (\Delta - B)(1-D)}$ are well-separated from the continuum edge $E = \Delta - B$ close to $\phi = \pi$.

We note that for stronger scattering, $1 - D \sim (\Delta - B)/\Delta$, the bound states and the continuum states contribute equally to the Josephson current near $\phi = \pi$, $I^{(1)}(\phi)\sim I^{(2)}(\phi)$.\\

\subsubsection{Approximate expression for the Josephson current in the topological phase, \texorpdfstring{$B > \Delta$}{B > Δ}}
\label{app:jc_aprx_top}

Here we employ Eq.~\eqref{eq:app_current_top} to obtain an approximate expression [Eq.~\eqref{eq:I_top}] for the Josephson current under the assumptions $|\pi + \phi_0^\mathrm{sc}-\phi| \ll  \phi_0^\mathrm{sc}$
$\Delta (1-D) \ll B - \Delta \ll \Delta$, $|\gamma|\ll 1$. We begin the calculation by analyzing the first term in Eq.~\eqref{eq:app_current_top}. To compute this contribution, we employ the approximate expression for $\Lambda(E,\phi)$ given by Eq.~\eqref{eq:app_lambda_top_simpl}. This expression can be further simplified in the limit we consider here. First, since $|\pi + \phi_0^\mathrm{sc}-\phi| \ll  \phi_0^\mathrm{sc}$ and $\phi_0^\mathrm{sc} \approx 2(B-\Delta)/\Delta$ we estimate $E_0 \approx (B - \Delta)$. Then, in virtue of inequality $\Delta(1-D)\ll B-\Delta$ the terms $\Delta^2(1-D)^2$ can be disregarded in comparison with $E_0^2$. As a result, we find
\begin{equation}
    \Lambda(i\mathcal{E},\phi) = 2\Delta(B-\Delta)\left(E_0^2+\mathcal{E}^2\right)\Bigl[1 + \Bigl(1+\frac{\mathcal{E}^2}{(B-\Delta)^2}\Bigr)^{\frac{1}{2}}\Bigr].
\end{equation}
Consequently,
\begin{equation}
\label{eq:app_p1}
    -\frac{e}{2\pi} \int_{-\infty}^{+\infty} d\mathcal{E} \partial_\phi \ln \Lambda(i\mathcal{E},\phi)\approx -e\Delta/2.
\end{equation}
Next, we account for the second term in Eq.~\eqref{eq:app_current_top}. In the considered phase interval it is given by $-2e \Theta (\pi + \phi_0^\mathrm{sc} - \phi) \partial_\phi E_0^\mathrm{sc}(\phi)$.
Combining this contribution with Eq.~\eqref{eq:app_p1} and using the approximate expression for $E_0^\mathrm{sc}(\phi)$, Eq.~\eqref{eq:boundsc}, we obtain Eq.~\eqref{eq:I_top}.

\end{document}